\documentclass[journal,english,twoscolumn,journal]{IEEEtran}
\usepackage[T1]{fontenc}
\usepackage{float}
\usepackage{amsthm}
\usepackage{amsmath}
\usepackage{amssymb}
\usepackage{stackrel}
\usepackage{graphicx}
\usepackage{indentfirst}
\usepackage{bm}
\usepackage{babel}

\pdfpageheight\paperheight
\pdfpagewidth\paperwidth

\floatstyle{ruled}
\newfloat{algorithm}{tbp}{loa}
\providecommand{\algorithmname}{Algorithm}
\floatname{algorithm}{\protect\algorithmname}

\theoremstyle{plain}
\newtheorem{thm}{\protect\theoremname}
\theoremstyle{remark}
\newtheorem{rem}{\protect\hskip 1em\remarkname}
\theoremstyle{plain}
\newtheorem{cor}{\protect\corollaryname}
\theoremstyle{plain}

\IEEEoverridecommandlockouts
\usepackage{ifpdf}
\usepackage{cite}
\usepackage{amsmath}
\usepackage{float}
\usepackage{array}
\usepackage{makecell}
\usepackage{caption2}
\usepackage{url}
\usepackage{accents}
\usepackage{statex}
\makeatletter
\def\wideubar{\underaccent{{\cc@style\underline{\mskip10mu}}}}
\def\Wideubar{\underaccent{{\cc@style\underline{\mskip8mu}}}}
\makeatother

\makeatletter
\def\widebar{\accentset{{\cc@style\underline{\mskip10mu}}}}
\def\Widebar{\accentset{{\cc@style\underline{\mskip8mu}}}}
\makeatother
\usepackage{graphicx}
\usepackage[outerbars,color]{changebar}
\cbcolor{red}

\makeatother

\usepackage{babel}
\providecommand{\corollaryname}{Corollary}
\providecommand{\remarkname}{Remark}
\providecommand{\theoremname}{Theorem}
\providecommand{\lemmaname}{Lemma}

\begin{document}

\title{Optimization of the Energy-Efficient Relay-Based massive IoT Network}

\author{Tiejun Lv, \emph{Senior Member, IEEE}, Zhipeng Lin, \emph{Student
Member, IEEE}, Pingmu Huang, and Jie Zeng, \emph{Senior Member, IEEE} \thanks{The financial support of the National Natural Science Foundation of
China (NSFC) (Grant No. 61671072) is gratefully acknowledged.

T. Lv, Z. Lin, P. Huang, and J. Zeng are with the School of Information and Communication
Engineering, Beijing University of Posts and Telecommunications, Beijing
100876, China (e-mail: \{lvtiejun, linlzp, pmhuang, zengjie\}@bupt.edu.cn).} 
}
\maketitle
\begin{abstract}
To meet the requirements of high energy efficiency (EE) and large
system capacity for the fifth-generation (5G) Internet of Things (IoT),
the use of massive multiple-input multiple-output (MIMO) technology
has been launched in the massive IoT (mIoT) network, where a large
number of devices are connected and scheduled simultaneously. This
paper considers the energy-efficient design of a multi-pair decode-and-forward
relay-based IoT network, in which multiple sources simultaneously
transmit their information to the corresponding destinations via a
relay equipped with a large array. In order to obtain an accurate
yet tractable expression of the EE, firstly, a closed-form expression
of the EE is derived under an idealized simplifying assumption, in
which the location of each device is known by the network. Then, an
exact integral-based expression of the EE is derived under the assumption
that the devices are randomly scattered following a uniform distribution
and transmit power of the relay is equally shared among the destination
devices. Furthermore, a simple yet efficient lower bound of the EE
is obtained. Based on this, finally, a low-complexity energy-efficient
resource allocation strategy of the mIoT network is proposed under
the specific quality-of-service (QoS) constraint. The proposed strategy
determines the near-optimal number of relay antennas, the near-optimal
transmit power at the relay and near-optimal density
of active mIoT device pairs in a given coverage area. Numerical results
demonstrate the accuracy of the performance analysis and the efficiency
of the proposed algorithms. 
\end{abstract}

\begin{IEEEkeywords}
Energy efficiency, resource allocation, massive MIMO, decode-and-forward
relay, green mIoT.
\end{IEEEkeywords}

\section{Introduction}

\global\long\def\figurename{Fig.}

Internet of Things (IoT), an emerging technology attracting significant
attention, promotes a heightened level of awareness about our world
and has been used in various areas, such as governments, industry,
and academia \cite{Stankovic1,add1}. In IoT, not only are various
things (e.g., sensor devices and cloud computing systems) with substantial
energy consumption, but also the connection of things (e.g., radio
frequency identification (RFID) and fifth-generation (5G) network)
and the interaction of things (e.g., data sensing and communications)
are consuming a large amount of energy. Improving energy efficiency
(EE) has become one of the main goals and design challenges for the
presented IoT networks \cite{energy}. In order to achieve the most
efficient energy usage, various innovative `\textit{green IoT}' techniques
have been developed during the last few years \cite{GreenIoT,GreenIoT2,J3}.

On the other hand, connectivity is the foundation for a IoT network.
It is envisioned that billions of devices will be connected in the
5G IoT network by 2020 to build a smart city \cite{MckinseyCompan,add2}.
As one major segment of the IoT network, the \textit{massive IoT (mIoT)}
refers to the applications that are enabled by connecting a large
number of IoT devices to an internet-enabled system \cite{JunZou,eeIot}.
This network is typically used for the scenarios characterized by
low power, wide coverage and strong support for devices on a massive
scale, such as agriculture production detection, power utilization
collection, medical monitoring and vehicle scheduling \cite{MWang,IoTbook,add3}.
There is a target that connection density in the urban environment
will be 1 million $\mathrm{devices/km^{2}}$ \cite{Americas,Qualcomm}.
Considering the connection target and the energy limitation of mIoT
networks, the massive multiple-input multiple-output (MIMO) technique
(equipped with a large-scale antenna array), which can increase the
network capacity 10 times or more without requiring more spectrum
and simultaneously improve the EE of wireless systems on the order
of 100 times \cite{MIMONOMA,massiveMIMO}, has attracted increasing
attention on utilization in mIoT networks \cite{MMIot,J2,J1}.
As presented in \cite{ngo_energy_2013}, as one of the major enabling technologies
for 5G wireless systems, massive MIMO systems are capable of increasing
the EE by orders of magnitude compared to single-antenna systems,
in particularly when combined with simultaneous scheduling of a large
number of terminals (e.g., tens or hundreds) \cite{lu2014overview,7446253}.
Relying on a realistic power consumption model, the authors of \cite{bjornson_optimal_2015}
proved that the optimal system parameters are capable of maximizing
the EE in multi-device massive MIMO systems. In \cite{joung_energy-efficient_2014},
low-complexity antenna selection methods and power allocation algorithms
were proposed to improve the EE of large-scale distributed antenna
systems. The authors of \cite{FP_Z2} intended to optimize the global
EE of the uplink and downlink of multi-cell massive MIMO. A joint
pilot assignment and resource allocation strategy was studied in \cite{nguyen2015resource}
to maximize the EE of multi-cell massive MIMO networks.

As a parallel research avenue, a relay-based mIoT network was shown
to constitute a promising technique of expanding the coverage, reducing
the power consumption and achieving energy-efficient transmission.
In \cite{GreenIoT2}, by optimizing the available bandwidth of a relay-based
mIoT network, the energy consumption of all the relay BSs is minimized.
Similar to the observations in single-hop massive MIMO systems, it
was shown in \cite{suraweera_multi-pair_2013} that by invoking a
relay equipped with a large-scale antenna array and a simple relay
transceiver (e.g., linear zero-forcing (ZF) transceiver), the spectrum
efficiency (SE) of a two-hop relay system becomes proportional to
the number of relay antennas. Therefore, the combination of massive
MIMO and cooperative relaying constitutes an appealing option for
energy-efficient mIoT networks.

When writing this paper, we find that the existing literature rarely
focused on the research of the relay-based mIoT networks and that
on massive MIMO aided relay systems mainly paid attention to the analysis
of the SE. For instance, the asymptotic SE of massive MIMO aided relay
systems was investigated in \cite{Jin_liang}, while the SE of massive
MIMO relay systems was studied in \cite{ngo_multipair_2014}. However,
to the best of our knowledge, there are a paucity of contributions
to the energy-efficient transmission and resource allocation strategies
in massive MIMO relay-based mIoT systems. It is challenging to extend
the existing energy-efficient designs conceived for single-hop massive
MIMO systems \cite{bjornson_optimal_2015,joung_energy-efficient_2014,nguyen2015resource}
to the ones adopted in massive MIMO relay systems. Due to the fact
that, compared to single-hop transmission schemes, both the design
of the signal processing schemes used at the relay and the performance
analysis of the massive MIMO relay systems are fundamentally dependent
on the more complex two-hop channels. Therefore, it is important to
design energy-efficient resource allocation strategy for a massive
MIMO relay-based mIoT network.

Contrary to the above background, in this paper, we consider the performance
analysis and energy-efficient resource allocation optimization of
a massive MIMO decode-and-forward (DF) relay based mIoT network supporting
multiple pairs of mobile mIoT devices. We assume that the channel
state information (CSI) is estimated relying on the minimum mean-square
error (MMSE) criterion, and the relay employs a low-complexity linear
ZF transceiver. 
The main contributions of this paper are summarized as follows.
\begin{itemize}
\item Assuming that the location of each mobile device is known, 
we derive a closed-form expression of the EE in the considered mIoT
network using a massive MIMO aided DF relay. Additionally, a simplified
analytical expression is also derived for the special case where equal
transmit power is allocated to all destination devices by the relay.
\item Furthermore, assuming that the device locations follow a uniform random
distribution, we derive an exact integral-based expression of the
EE. However, since this expression cannot be integrated out, a simple
yet efficient lower bound of the EE is also derived. The analytical
results lay the foundation of predicting the EE and of understanding
how it changes with respect to the transmit power, the number of relay
antennas and the number of active mIoT device pairs.
\item Based on the lower bound derived, we propose an energy-efficient resource
allocation strategy which determines the near-optimal relay transmit
power, the near-optimal number of relay antennas and the near-optimal
number of active mIoT device pairs under the given quality-of-service
(QoS) constraint. The original EE optimization problem relying on
the exact but intractable EE expression is transformed into a problem
that maximizes the lower bound of the EE. This transformation makes
it feasible for us to solve the latter EE optimization problem, which
eventually gives the near-optimal system configuration of energy-efficient
massive MIMO relay systems supporting multiple mIoT device pairs.
\end{itemize}
The remainder of this paper is organized as follows. Both the system
model and transmission scheme of the mIoT network using the massive
MIMO aided multi-pair DF relay are described in Section II. In Section
III, the EE optimization problem is formulated by employing a realistic
power consumption model. In Section IV, we derive the EE expressions
with known device locations and uniform random distribution of device
locations, respectively. Then, in Section V, an energy-efficient resource
allocation strategy is proposed based on the EE expressions derived.
Numerical results are provided under diverse system configurations.
In Section VII, we analyze the convergence and the computational complexity
of the proposed algorithms. Finally, our conclusions are drawn in
Section VIII.

\emph{Notations:} We use uppercase and lowercase boldface letters
for denoting matrices and vectors, respectively. $(\cdot)^{H}$, $(\cdot)^{T}$
and $(\cdot)^{\dagger}$ denote the conjugate transpose, transpose
and pseudo-inverse, respectively. $||\cdot||$, $\textnormal{tr}(\cdot)$,
$\mathbb{E}[\cdot]$ and $\mathbb{V}ar[\cdot]$ stand for the Euclidean
norm, the trace of matrices, the expectation and variance operators,
respectively. $[\mathbf{A}]_{i,j}$ represents the entry at the $i$-th
row and the $j$-th column of a matrix $\mathbf{A}$. $\mathcal{CN}(\mathbf{0},\mathbf{\Theta})$
denotes the circularly symmetric complex Gaussian distribution with
zero mean and the covariance matrix $\mathbf{\Theta}$, while $\xrightarrow{a.s.}$
denotes the \emph{almost sure} convergence. $_{2}F_{1}(\cdot)$ represents
the hypergeometric function, and $|\mathcal{A}|$ denotes the cardinality
of a set $\mathcal{A}$. Finally, $\left[\cdot\right]^{+}$ denotes
$\max\left\{ 0,\cdot\right\} $.

\section{System Model and Transmission Scheme\label{sec:System-Model}}

\begin{figure}[t]
\begin{centering}
\includegraphics[width=7cm]{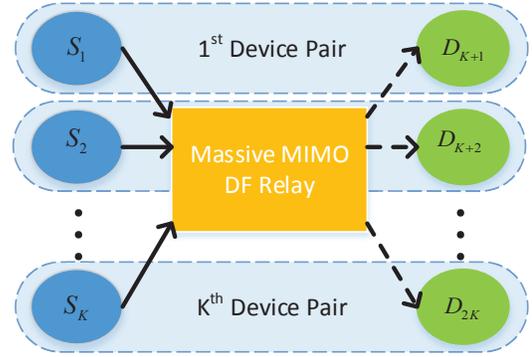}
\par\end{centering}
\protect\caption{The system model considered: A two-hop relay system supports $K$
active mIoT devices communicating with their individual destination
mIoT devices via a massive MIMO aided DF relay, which mitigates the
inter-stream interference with a large number of antennas.}
\label{sys_fig}
\end{figure}

\begin{figure}[tbh]
\begin{centering}
\includegraphics[width=8cm]{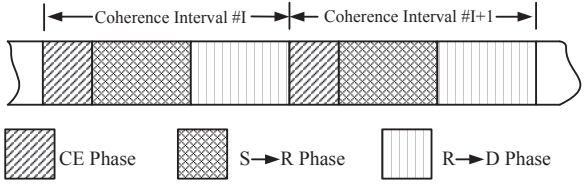}
\par\end{centering}
\protect\caption{Partitioning of a coherence interval.}
\label{division_coherence_interval}
\end{figure}

As shown in Fig. \ref{sys_fig}, we consider a massive MIMO aided
dual-hop DF relay mIoT system supporting $K$ pairs of single-antenna
source-destination mIoT devices (i.e., active mIoT device pairs),
which are selected from $N$ pairs of candidate mIoT devices for data
transmission with the aid of the $M$-antenna ($K\ll M$) relay\footnote{It was shown in \cite{interference} that by using simple relay transceivers
(e.g., linear ZF-based transceivers), a massive MIMO relay system
is capable of significantly alleviating the interference among different
data streams. }. The system operates over a bandwidth of $B$ Hz and the channels
are static within the time/frequency coherence interval of $T=B_{C}T_{C}$
symbol duration, where $B_{C}$ and $T_{C}$ are the coherence bandwidth
and coherence time, respectively. Particularly, each mIoT device as
well as the relay uses the total bandwidth of $B$ Hz. We focus on
the active mIoT device pairs and assume that the $k$-th mIoT device
(source node) demands to communicate with the $\left(k+K\right)$-th
device (destination node)\footnote{In this paper, the direct link between any source node and destination
node is ignored due to large path-loss.}. The set of active mIoT device pairs is denoted by ${\cal S}$, satisfying
$|{\cal S}|=K$. The relay operates in the half-duplex time-division
duplexing (TDD) mode. Each coherence interval is divided into three
time phases, as shown in Fig. \ref{division_coherence_interval},
namely the channel estimation (CE) phase, the source-to-relay transmission
(${\rm S}\to{\rm R}$) phase, and the relay-to-destination transmission
(${\rm R}\to{\rm D}$) phase.

Let ${{\bf G}_{{\rm S}}}=\left[{\bf g}_{{\rm S},1},\cdots,{\bf g}_{{\rm S},K}\right]\in\mathbb{C}^{M\times K}$
and ${\bf G}_{{\rm D}}=\left[{\bf g}_{{\rm D},1},\cdots,{\bf g}_{{\rm D},K}\right]^{T}\in\mathbb{C}^{K\times M}$
denote the channel matrices from the $K$ active sources to the relay
and from the relay to the $K$ active destinations, respectively.
The channel matrices characterize both the small-scale fading (SSF)
and the large-scale fading (LSF). More precisely, ${\bf {G}_{\text{S}}}$
and ${\bf {G}_{\text{D}}}$ can be expressed as
\begin{align}
\mathbf{G}_{\text{S}}=\mathbf{H}_{\text{S}}\mathbf{D}_{\text{S}}^{1/2},\ {\bf G}_{\text{D}} & ={\bf D}_{\text{D}}^{1/2}{\bf H}_{\text{D}},\label{eq:1}
\end{align}
where $\mathbf{H}_{\text{S}}\in\mathbb{C}^{M\times K}$ and $\mathbf{H}_{\text{D}}\in\mathbb{C}^{K\times M}$
are the SSF channel matrices and their entries are independent and
identically distributed (i.i.d.) with ${\cal CN}\left(0,1\right)$.
The LSF channel matrices $\mathbf{D}_{\text{S}}$ and $\mathbf{D}_{\text{D}}$
are modelled as diagonal matrices with $\left[\mathbf{D}_{\text{S}}\right]_{k,k}=\beta_{k}$
and $\left[\mathbf{D}_{\text{D}}\right]_{k,k}=\beta_{k+K}$, $k=1,2,\cdots,K$,
respectively.

\subsection{CE at the Relay}

In the CE phase, the relay acquires the CSI of active devices by using
the MMSE channel estimator given in \cite{MMSE1995}. Let $\hat{{\bf G}}_{{\rm S}}$
and $\hat{{\bf G}}_{{\rm D}}$ be the channel estimates of ${\bf G}_{{\rm S}}$
and ${\bf G}_{{\rm D}}$, respectively. Then, we have
\begin{align}
{\bf G}_{{\rm S}} & =\hat{{\bf G}}_{{\rm S}}+{\tilde{{\bf G}}}_{{\rm S}},\ {\bf G}_{{\rm D}}=\hat{{\bf G}}_{{\rm D}}+\tilde{{\bf G}}_{{\rm D}},\label{channel_model}
\end{align}
where ${\tilde{{\bf G}}}_{{\rm S}}$ and ${\tilde{{\bf G}}}_{{\rm D}}$
are the complex-valued Gaussian distributed estimation error matrices
of ${\bf G}_{{\rm S}}$ and ${\bf G}_{{\rm D}}$, respectively. According
to the orthogonality principle \cite{MMSE1995}, $\hat{{\bf G}}_{{\rm S}}$
and $\tilde{{\bf G}}_{{\rm S}}$ are independent of each other. Similarly,
$\hat{{\bf G}}_{{\rm D}}$ and $\tilde{{\bf G}}_{{\rm D}}$ are independent
of each other as well.

For the clarity of analysis, we temporarily assume that the LSF channel
matrices, $\mathbf{D}_{\text{S}}$ and $\mathbf{D}_{\text{D}}$, are
perfectly estimated. In Section IV-B and in its subsequent sections,
this assumption will then be removed. According to \eqref{channel_model},
the columns of $\hat{{\bf G}}_{{\rm S}}$, $\tilde{{\bf G}}_{{\rm S}}$,
$\hat{{\bf G}}_{{\rm D}}$ and $\tilde{{\bf G}}_{{\rm D}}$ obey the
distributions of
\begin{equation}
\begin{split}\hat{{\bf g}}_{{\rm S},k} & \sim{\cal CN}\left({\bf 0},\beta_{k}^{'}{\bf I}_{M}\right),\\
\hat{{\bf g}}_{{\rm S},k} & \sim{\cal CN}\left({\bf 0},\beta_{k}-\beta_{k}^{'}{\bf I}_{M}\right),\\
\hat{{\bf g}}_{{\rm D},k} & \sim{\cal CN}\left({\bf 0},\beta_{k+K}^{'}{\bf I}_{M}\right),\\
\hat{{\bf g}}_{{\rm D},k} & \sim{\cal CN}\left({\bf 0},\beta_{k+K}-\beta_{k+K}^{'}{\bf I}_{M}\right),
\end{split}
\end{equation}
where $\beta_{i}^{'}=\frac{\tau_{{\rm r}}\rho_{{\rm p}}\beta_{i}^{2}}{1+\tau_{{\rm r}}\rho_{{\rm p}}\beta_{i}}$,
$i=1,2,\cdots,2K$. Furthermore, $\rho_{{\rm p}}$ is the ratio of
the transmit power of each pilot symbol to the noise power at the
relay's receiver, while $\tau_{{\rm r}}$ ($\tau_{{\rm r}}\geq2K$)
is the pilot sequence length of each device. 

\subsection{Data Transmission}

The relay uses the channel estimates obtained above and the low-complexity
linear ZF transceivers. More specifically, the relay uses a ZF receiver
to detect the signals transmitted from the $K$ active sources, and
then it uses a ZF transmit precoding scheme to forward the signals
to the $K$ active destinations.

In the $\text{S}\to\text{R}$ phase, $K$ sources simultaneously transmit
their signals $\mathbf{x}_{\text{d}}=\sqrt{P_{{\rm tx,d}}}\mathbf{s}\in\mathbb{C}^{K\times1}$
to the relay, in which $\mathbf{s}=[s_{1},\dots,s_{k},\dots,s_{K}]^{T}$
is the information-bearing symbol vector satisfying $\mathbb{E}[\mathbf{s}\mathbf{s}^{H}]={\bf I}_{K}$,
$s_{k}$ is the symbol delivered from the $k$-th mIoT device to the
relay, and $P_{{\rm tx,d}}$ is the average transmit power of each
mIoT device. The signal $\mathbf{y}_{\text{R}}\in\mathbb{C}^{M\times1}$
received at the relay is given by
\begin{align}
\mathbf{y}_{\text{R}}=\mathbf{G}_{\text{S}}\mathbf{x}_{\text{d}}+\mathbf{n}_{\text{R}},
\end{align}
where $\mathbf{n}_{\text{R}}\in\mathbb{C}^{M\times1}$ denotes the
additive white Gaussian noise (AWGN) obeying $\mathcal{CN}({\bf 0},\sigma_{{\rm R}}^{2}\mathbf{I}_{M})$
at the relay.

The relay performs ZF detection relying on ${\bf y_{\text{R}}}$.
More specifically, upon multiplying with the ZF filtering matrix ${\mathbf{F}}_{1}=\left(\hat{{\bf G}}_{{\rm S}}^{H}\hat{{\bf G}}_{{\rm S}}\right)^{-1}\hat{{\bf G}}_{{\rm S}}^{H}\in{\mathbb{C}}^{K\times M}$,
the transmitted symbols having been superimposed over the channel
are separated into $K$ non-interfering symbols, which are denoted
as ${\bf x}_{\text{R}}\in\mathbb{C}^{K\times1}$ and given by
\begin{align}
{\bf x}_{\text{R}} & ={{\bf F}_{\text{1}}}{\bf y}_{\text{R}}={\bf F}_{{\rm 1}}{\bf G}_{{\rm S}}{\bf x}_{\text{d}}+{\bf F}_{{\rm 1}}{\bf n}_{{\rm R}}.
\end{align}

Therefore, the $k$-th symbol $x_{\text{R},k}$ (i.e., the $k$-th
element of ${\bf x}_{\text{R}}$) is given by 
\begin{align}
\begin{split}x_{\text{R},k} & =\sqrt{P_{{\rm tx,d}}}{\mathbb{E}}\left[{\bf f}_{1,k}^{H}{\bf g}_{{\rm S},k}\right]s_{k}+\sqrt{P_{{\rm tx,U}}}\sum_{j\neq k}^{K}{\bf f}_{1,k}^{H}{\bf g}_{{\rm S},j}{s}_{j}\\
 & +\sqrt{P_{{\rm tx,d}}}\left({\bf f}_{1,k}^{H}{\bf g}_{{\rm S},k}-{\mathbb{E}}\left[{\bf f}_{1,k}^{H}{\bf g}_{{\rm S},k}\right]\right)s_{k}+{\bf f}_{1,k}^{H}{\bf n}_{\text{R}},\label{RE_{R}}
\end{split}
\end{align}
where ${\bf f}_{1,k}$ is the $k$-th column of $\mathbf{F}_{1}$.
Then, the signal-to-interference-plus-noise ratio (SINR) of the $k$-th
device pair in the $\text{S}\to\text{R}$ phase is formulated as\cite{R2aaa}

\begin{align}
\begin{split}\gamma_{k}^{(1)}=P_{{\rm tx,U}}\begin{vmatrix}{\mathbb{E}}\left[{\bf f}_{1,k}^{H}{\bf g}_{{\rm S},k}\right]\end{vmatrix}^{2}/(P_{{\rm tx,d}}{\mathbb{V}}{\rm ar}\left({\bf f}_{1,k}^{H}{\bf g}_{{\rm S},k}\right)\\
+P_{{\rm tx,d}}\sum_{j\neq k}^{K}{\mathbb{E}}\left[\begin{vmatrix}{\bf f}_{1,k}^{H}{\bf g}_{{\rm S},j}\end{vmatrix}^{2}\right]+\sigma_{{\rm R}}^{2}{\mathbb{E}}\left[\begin{Vmatrix}{\bf f}_{1,k}\end{Vmatrix}^{2}\right]).\label{SNR1}
\end{split}
\end{align}

During the $\text{R}\to\text{D}$ phase, the relay decodes the information
symbol vector ${\bf s}$ from ${\bf x}_{\textrm{R}}$ as $\hat{{\bf s}}=\left[\hat{s}_{1},\cdots,\hat{s}_{k},\cdots,\hat{s}_{K}\right]^{T}$
that satisfies $\mathbb{E}[{\hat{\mathbf{s}}}{\hat{\mathbf{s}}}^{H}]={\bf I}_{K}$,
and multiplies the ZF precoding matrix of ${\bf F}_{2}=\hat{{\bf G}}_{{\rm D}}^{H}\left(\hat{{\bf G}}_{{\rm D}}\hat{{\bf G}}_{{\rm D}}^{H}\right)^{-1}\in{\mathbb{C}}^{M\times K}$
both by the power allocation matrix at the relay and by $\hat{{\bf s}}$,
yielding
\begin{align}
\underline{{\bf x}}_{\text{R}}={\bf F}_{\text{2}}{\bf P}\hat{{\bf s}},\label{eq:-7}
\end{align}
which is broadcast to all the $K$ active destinations. Here, ${\bf P}=\text{diag}\left\{ \sqrt{p_{1}},\dots,\sqrt{p_{K}}\right\} $
is the power allocation matrix used at the relay. The long-term average
total transmit power constraint at the relay is thus given by
\begin{align}
P_{{\rm tx,R}}\triangleq{\rm tr}\left(\underline{{\bf x}}_{\text{R}}\underline{{\bf x}}_{\text{R}}^{H}\right)\leq P_{{\rm R\max}},\label{RPfactor}
\end{align}
where $P_{{\rm tx,R}}$ represents the average total transmit power
at the relay, and $P_{{\rm R\max}}$ is the maximum average total
transmit power available at the relay. 
The signal ${\bf y}_{\text{D}}\in\mathbb{C}^{K\times1}$ received
at the $K$ active destinations is given by
\begin{align}
{\bf y}_{\text{D}}={\bf G}_{\text{D}}\underline{{\bf x}}_{\text{R}}+{\bf n}_{\text{D}}={\bf P}\hat{{\bf s}}+{\tilde{{\bf G}}}_{{\rm D}}\underline{{\bf x}}_{\text{R}}+{\bf n}_{\text{D}},
\end{align}
where $\mathbf{n}_{\text{D}}\in\mathbb{C}^{K\times1}$ denotes the
AWGN at the destinations and it obeys $\mathcal{CN}({\bf 0},\sigma_{{\rm D}}^{2}\mathbf{I}_{K})$.
Thus, the signal received at the $k$-th active destination is 
\begin{align}
y_{\text{D},k} & =\sqrt{p_{k}}{\mathbb{E}}\left[{\bf g}_{{\rm {D}},k}^{H}{\bf f}_{2,k}\right]\hat{s}_{k}+\sqrt{p_{k}}\left({\bf g}_{{\rm {D}},k}^{H}{\bf f}_{2,k}-{\mathbb{E}}\left[{\bf g}_{{\rm {D}},k}^{H}{\bf f}_{2,k}\right]\right)\hat{s}_{k}\nonumber \\
 & +\sum_{j\neq k}^{K}\sqrt{p_{j}}{\bf g}_{\text{D},k}^{H}{\bf f}_{2,j}\hat{s}_{j}+n_{\text{D},k},\label{RS_D}
\end{align}
where $n_{{\rm D},k}$ is the $k$-th element of ${\bf n}_{{\rm D}}$,
while ${\bf f}_{2,k}$ is the $k$-th column of ${\bf F}_{\text{2}}$.
The SINR at the destination $U_{k+K}$ for the $k$-th transmitted
data stream is given by\cite{R2aaa} 
\begin{align}
\gamma_{k}^{(2)}=\frac{p_{k}\begin{vmatrix}{\mathbb{E}}\left[{\bf g}_{{\rm {D}},k}^{H}{\bf f}_{2,k}\right]\end{vmatrix}^{2}}{p_{k}{\mathbb{V}}{\rm ar}\left({\bf g}_{{\rm {D}},k}^{H}{\bf f}_{2,k}\right)+\sum_{j\neq k}^{K}p_{j}{\mathbb{E}}\left[\begin{vmatrix}{\bf g}_{{\rm D},k}^{H}{\bf f}_{2,j}\end{vmatrix}^{2}\right]+\sigma_{{\rm D}}^{2}}.\label{SNR2}
\end{align}

As a result, the end-to-end achievable rate of the $k$-th mIoT device
pair can be given by
\begin{align}
{\cal R}_{k}=\min\left\{ {\cal R}_{k}^{\left(1\right)},\ {\cal R}_{k}^{\left(2\right)}\right\} ,
\label{ar_e2e}
\end{align}
where ${\cal R}_{k}^{\left(1\right)}=\log_{2}\left(1+\gamma_{k}^{\left(1\right)}\right)$,
${\cal R}_{k}^{\left(2\right)}=\log_{2}\left(1+\gamma_{k}^{\left(2\right)}\right)$.

An overhead occupying $\tau_{{\rm r}}=2K$ symbol intervals is used
to facilitate the pilot-based CE during each coherence interval of
$T$ symbols. Therefore, we have to take the overhead-induced dimensionality
loss of $2K/T$ into account when calculating the sum rate which is
expressed as
\begin{align}
\mathcal{R}_{{\rm sum}}=\left(1-\frac{2K}{T}\right)\frac{B}{2}\sum_{k=1}^{K}{\cal R}_{k}.\label{SE}
\end{align}


\section{EE Optimization Problem Formulation}

We employ a realistic power consumption model similar to those used
in \cite{bjornson_optimal_2015,joung_energy-efficient_2014,joung_ema_2014}.
The total power consumption of the system considered is quantified
as
\begin{align}
{P_{{\rm tot}}}={P}_{{\rm PA}}+{P}_{{\rm C}},\label{power_C}
\end{align}
where we have
\begin{align}
{P}_{{\rm PA}} & =\frac{\eta_{{\rm PA,d}}^{-1}}{2}\left(1-\frac{2K}{T}\right)KP_{{\rm tx,d}}+\frac{\eta_{{\rm PA,d}}^{-1}4K^{2}}{T}\rho_{{\rm r}}\sigma_{{\rm r}}^{2}\nonumber \\
 & +\frac{\eta_{{\rm PA,R}}^{-1}}{2}\left(1-\frac{2K}{T}\right)P_{{\rm tx,R}},\\
{P}_{{\rm C}} & =P_{{\rm FIX}}+P_{{\rm TC}}+P_{{\rm SIG}},
\end{align}
with ${P}_{{\rm PA}}$ representing the power consumed by power amplifiers
(PAs), in which $P_{{\rm tx,d}}$ and $\rho_{{\rm r}}\sigma_{{\rm r}}^{2}$
are the data transmit power and pilot transmit power of each active
source device, respectively, while $\eta_{{\rm PA,d}}\in(0,1)$ and
$\eta_{{\rm PA,R}}\in(0,1)$ are the efficiency of the PAs at the
devices and at the relay, respectively. Furthermore, ${P}_{{\rm C}}$
denotes the total circuit power consumption, in which $P_{{\rm FIX}}$
is a constant accounting for the fixed power consumption required
for control signalling, site-cooling and the load-independent base-band
signal processing, $P_{{\rm TC}}$ indicates the power consumption
of the transceiver's radio-frequency (RF) chains, and $P_{{\rm SIG}}$
accounts for the power consumption of the load-dependent signal processing.
To be more specific, we have
\begin{align}
\begin{split}P_{{\rm TC}} & =MP_{{\rm R}}+2KP_{{\rm d}}+P_{{\rm SYN}},\\
P_{{\rm SIG}} & =\frac{B}{T}\frac{8MK^{2}}{L_{{\rm R}}}+B\left(1-\frac{2K}{T}\right)\frac{4MK}{L_{{\rm R}}}\\
 & +\frac{B}{T}\frac{1}{3L_{{\rm R}}}(K^{3}+9MK^{2}+3MK),
\end{split}
\label{PC_2}
\end{align}
where $P_{{\rm R}}$ and $P_{{\rm d}}$ represent the power required
to run the circuit components attached to each antenna at the relay
and at the devices, respectively, while $P_{{\rm SYN}}$ is the power
consumed by the oscillator. The first term of $P_{{\rm SIG}}$ describes
the power consumption of CE, while the remaining two terms account
for the power required by the computation of the ZF detection matrix
${\bf F}_{1}$ and the ZF precoding matrix ${\bf F}_{2}$. Still referring
to \eqref{PC_2}, $L_{{\rm R}}$ denotes the computational efficiency
quantified in terms of the complex-valued arithmetic operations per
Joule at the relay. As a result, the EE $\eta_{{\rm EE}}$ {[}bits/Joule{]}
is defined as\footnote{It can be readily seen from \eqref{power_C}-\eqref{PC_2} that the
total power consumption highly depends on the number of relay antennas
$M$, on the relay power allocation matrix ${\bf P}$ and on the selection
of the active mIoT device pairs ${\cal S}$. Choosing an appropriate
power consumption model is of paramount importance, when dealing with
the energy-efficient resource allocation in massive MIMO aided multi-pair
relay mIoT.}
\begin{align}
\eta_{{\rm EE}}\left(\mathbf{P},\mathcal{S},M\right) & =\mathcal{R}_{{\rm sum}}/P_{{\rm tot}}.\label{Sim_EE}
\end{align}

It is plausible that $\eta_{{\rm EE}}$ is a function of the following
system resources: the power allocation matrix ${\bf P}$ used at the
relay, the set ${\cal S}$ of active mIoT device pairs, and the number
of the relay antennas, $M$. In this paper, the energy-efficient resource
allocation is formulated as the following optimization problem.
\begin{align}
\max_{\mathbf{P},\mathcal{S},M} & \quad\eta_{{\rm EE}}\left(\mathbf{P},\mathcal{S},M\right),\nonumber \\
\mathrm{s.t.} & \quad{\rm C1:}\ P_{{\rm tx,R}}\leq P_{{\rm R\max}},\nonumber \\
 & \quad{\rm C2}:\ \mathcal{S}\in\mathbb{U},\nonumber \\
 & \quad{\rm C3}:\ M\in\{1,2,\dots,M_{\max}\},\label{eq_ee_1}\\
 & \quad{\rm C4}:\ p_{k}\geq0,\ k=1,\cdots,\ K,\nonumber \\
 & \quad{\rm C5}:\ {\cal R}_{k}\geq{R}_{0},\ k=1,\cdots,\ K,\nonumber
\end{align}
where the objective function $\eta_{{\rm EE}}\left(\mathbf{P},\mathcal{S},M\right)$
is defined by \eqref{Sim_EE}. In \eqref{eq_ee_1}, ${\rm C1}$ ensures
that the sum of power allocated to the $K$ data streams does not
exceed the maximum transmit power available to the relay, while ${\rm C}2$
is a combinatorial constraint imposed on the device-pair selection,
where $\mathbb{U}$ denotes the group of all the available sets of
active device pairs. The constraint associated with the number of
relay antennas $M$ is specified by ${\rm C}3$, in which $M_{\max}$
is the largest possible number, and ${\rm C}4$ is the boundary constraint
of the relay power allocation variables. In addition, for some mIoT
application scenarios (e.g., agriculture production detection), system
energy is limited or electricity is generated by means of unpredictable
renewable energy, such as wind and solar. One requirement in terms
of the system operation is to guarantee appropriate QoS, which is
a profile associated to each data instance. QoS can regulate the nonfunctional
properties of information \cite{add4}. ${\rm C}5$ represents the
QoS constraint for each device pair, where ${\cal R}_{k}$ is the
achievable rate of the $k$-th pair device, and ${R}_{0}$ is a given
threshold.

\section{EE Analysis of the massive MIMO aided multi-pair DF relay system}

From \eqref{SE}, \eqref{Sim_EE} and \eqref{eq_ee_1}, we can see
that it is challenging to calculate the EE in real time, because the
EE depends on specific LSF channel coefficients and requires challenging
optimization involving matrix variables. To overcome this predicament,
in this section, we firstly derive a closed-form expression of the
EE under the assumption that the mIoT device locations are known.
Subsequently, an EE expression is provided for the scenario where
the device locations are assumed to be independent and uniformly distributed (i.u.d.) random variables in the coverage area. In these expressions, both the instantaneous
channel coefficients and the matrix variables will disappear, hence
the instantaneous CSI and the complex matrix calculations are no longer
needed in our resource allocation.

\subsection{EE Analysis Assuming Known Device Locations}

In this subsection, upon assuming that the mIoT device locations are
known \textit{a priori} (i.e., the LSF channel coefficients are assumed
to be perfectly estimated), we derive a closed-form expression of
the EE. As the system rate ${\cal R}_{k}$ for finite system dimensions
is difficult to calculate, we consider the large system limit, where
$M$ and $K$ grow infinitely large while keeping a finite ratio $M/K$.
However, as we use the asymptotic analysis only as a tool provide
tight approximations for finite $M,\ K$. In what follows, we will
derive deterministic approximations of the system rates ${\cal R}_{k}$.
Thus, considering both the SE and total power consumption, a closed-form
deterministic approximations of the EE in the system considered is
presented in the following theorem. 
\begin{thm}
Using linear ZF transceivers with imperfect CSI, as well as assuming
that the mIoT device locations are known a priori, the EE of the massive
MIMO aided multi-pair relay system considered can be calculated by
\begin{align}
{\eta}_{{\rm EE}}\left({\bf P},{\cal S},M\right) & \xrightarrow{a.s.}\frac{\left(1-\frac{2K}{T}\right){\frac{B}{2}{\cal R}}\left({\bf P},{\cal S},M\right)}{P_{{\rm tot}}\left({\bf P},{\cal S},M\right)},\label{AEE}
\end{align}
where
\begin{align}
{\cal R}\left({\bf P},{\cal S},M\right)=\sum_{k=1}^{K}\min\left\{ {\cal R}_{k}^{(1)},\ {\cal R}_{k}^{(2)}\right\}
\end{align}
with ${\cal R}_{k}^{(1)}$ and ${\cal R}_{k}^{(2)}$ being calculated
as
\begin{align}
\begin{split}{\cal R}_{k}^{(1)}\xrightarrow{a.s.}\log_{2}\left(1+\frac{\left(M-K\right)P_{{\rm tx,U}}\beta_{k}^{'}}{P_{{\rm tx,d}}A_{1}+\sigma_{{\rm R}}^{2}}\right),\\
{\cal R}_{k}^{(2)}\xrightarrow{a.s.}\log_{2}\left(1+\frac{p_{k}}{{\tilde{\beta}}_{k+K}P_{{\rm tx,R}}+\sigma_{{\rm D}}^{2}}\right),\label{ANSR1}
\end{split}
\end{align}
in which
\begin{align}
A_{1}=\sum_{j=1}^{K}\left(\beta_{j}-\beta_{j}^{'}\right)=\sum_{j=1}^{K}{\tilde{\beta}}_{j},\ {\tilde{\beta}}_{k+K}=\beta_{k+K}-\beta_{k+K}^{'}.\label{BB2}
\end{align}

The total power consumption ${P_{{\rm tot}}}\left({\bf P},{\cal S},M\right)$
is given by \eqref{power_C}, where we have
\begin{align}
P_{{\rm tx,R}} & \xrightarrow{a.s.}\frac{\sum_{k=1}^{K}p_{k}\left(\beta_{k+K}^{'}\right)^{-1}}{M-K}.\label{Ptx_R}
\end{align}
\end{thm}

\begin{IEEEproof}
See Appendix I.
\end{IEEEproof}
According to \eqref{Ptx_R}, the long-term average total transmit
power constraint \eqref{RPfactor} of the relay can be rewritten as
\begin{align}
\frac{\sum_{k=1}^{K}p_{k}\left(\beta_{k+K}^{'}\right)^{-1}}{M-K}\leq P_{{\rm R\max}}.\label{eq:power factor}
\end{align}
\begin{rem}
\noindent In Theorem 1, we obtain the closed-form expression of the
EE, which only depends on the LSF channel coefficients of active mIoT
device pairs and on the configurable system parameters. This expression
is a fundamental one that characterizes the relationship between the
EE and $\left({\bf P},{\cal S},M\right)$ for the general case, and
acts as source in the subsequent section. In the expression, the complicated
calculation involving large-dimensional matrix variables that represent
the SSF channel coefficients is avoided.
\end{rem}
In practical relay aided systems, the computational resources of the
relay are limited. Hence, optimally solving the mixed-integer nonlinear
optimization problem of \eqref{eq_ee_1} may become computationally
unaffordable to the relay. As shown in Fig. \ref{iter}, the average
EE performance of the brute-force search aided optimal power allocation
is only slightly higher than that of the equal power allocation strategy.
Therefore, the equal power allocation strategy can be used at the
relay for reducing the computational complexity of directly solving
\eqref{eq_ee_1}. More specifically, upon considering the equal power
allocation that satisfies \eqref{eq:power factor} for any $k=1,\dots,K$,
the power allocation coefficients are calculated as
\begin{align}
p_{k}=\frac{\left(M-K\right)P_{{\rm tx,R}}}{{A}_{2}},\ \forall k,\label{eq:Ave_power}
\end{align}
where ${A}_{2}=\sum_{k=1}^{K}\left(\beta_{k+K}^{'}\right)^{-1}$.
As a result, compared to the optimal power allocation \eqref{AEE}
that optimizes $p_{k}$ ($k=1,\cdots,K$) for each mIoT device pair,
it becomes feasible for us to only optimize the total transmit power
$P_{{\rm tx,R}}$, when the relay's transmit power is equally allocated
to all the destination devices.

Substituting \eqref{eq:Ave_power} into \eqref{AEE}, we arrive at
the following corollary concerning the EE under the assumption of
using equal power allocation at the relay.
\begin{cor}
The EE ${\widebar{\eta}}_{{\rm EE}}$ associated with the equal power
allocation at the relay is calculated as
\begin{align}
{\widebar{\eta}}_{{\rm EE}}\left(P_{{\rm tx,R}},{\cal S},M\right) & \xrightarrow{a.s.}\frac{\left(1-\frac{2K}{T}\right)\frac{B}{2}{\widebar{\cal R}}\left(P_{{\rm tx,R}},{\cal S},M\right)}{P_{{\rm tot}}\left(P_{{\rm tx,R}},{\cal S},M\right)},\label{APP1_EE}
\end{align}
where
\begin{align}
{\widebar{\cal R}}\left(P_{{\rm tx,R}},{\cal S},M\right)=\sum_{k=1}^{K}\min\left\{ {\widebar{\cal R}}_{k}^{(1)},{\widebar{\cal R}}_{k}^{(2)}\right\} ,\label{ar1}
\end{align}
with ${\widebar{\cal R}}_{k}^{(1)}={\cal R}_{k}^{(1)}$ (given by
\eqref{ANSR1}) and
\begin{align}
{\widebar{\cal R}}_{k}^{(2)}\xrightarrow{a.s.}\log_{2}\left(1+\frac{\left(M-K\right)P_{{\rm tx,R}}}{\left(\tilde{\beta}_{k+K}P_{{\rm tx,R}}+\sigma_{{\rm D}}^{2}\right)A_{2}}\right).\label{ASNR22}
\end{align}
Again, ${P_{{\rm tot}}}\left(P_{{\rm tx,R}},{\cal S},M\right)$ is
given by \eqref{power_C}.
\end{cor}

\subsection{EE Analysis Assuming i.u.d. Device Locations}

In the previous subsection, we have derived the closed-form expression
of the EE under the assumption that the device locations are known
\textit{a priori}. This assumption imposes an extremely high complexity
burden and implementation cost, especially in high-mobility environments,
because the channel coefficients will change rapidly and it is difficult
to select the active device pairs instantly in practical mobile communication
systems \cite{Lv_DOA}. In this subsection, we consider a more general
scenario in which the mIoT devices are assumed to be i.u.d. in the
relay's coverage area, and derive the corresponding EE expression
as a function of the total relay transmit power $P_{{\rm tx,R}}$,
the number of active mIoT device pairs $|{\cal S}|=K$ and the number
of relay antennas $M$. The EE expression obtained in this scenario
provides further insights into the selection of EE-optimal system
parameters.

We assume that the relay's coverage area is modelled as a disc and
the relay is located at the geometric center of this disc. Furthermore,
all the active source and destination devices are assumed to be i.u.d.
in the disc, whose radius $R$ satisfies $R_{\min}\leq R\leq R_{\max}$.
The LSF channel coefficient of the $k$-th active mIoT device is modelled
as $\beta_{k}=cl_{k}^{-\alpha}$, where $l_{k}$ is the distance between
the $k$-th mIoT device and the relay, $\alpha$ is the path-loss
exponent, and $c$ is the path-loss at the reference distance $R_{\min}$.
The probability density function (PDF) of $l_{k}$ is
\begin{align}
f\left(l_{k}\right) & =\frac{2l_{k}}{R_{\max}^{2}-R_{\min}^{2}},\ R_{\min}\leq l_{k}\leq R_{\max},
\end{align}
where $R_{\max}$ is the radius of the circular cell.

In Theorem 2, we first give the expression of the EE assuming i.u.d.
device locations.
\begin{thm}
Given the other parameters, using linear ZF transceivers with imperfect
SSF channel coefficients estimated by the MMSE estimator and assuming
that all the devices are i.u.d. in the relay's coverage area, the
EE of the massive MIMO aided multi-pair relay system considered with
equal relay power allocation is formulated as
\begin{align}
\widetilde{\eta}_{{\rm EE}}\left(P_{{\rm tx,R}},\ K,\ M\right) & \xrightarrow{a.s.}\frac{\left(1-\frac{2K}{T}\right)\frac{B}{2}\widetilde{{\cal R}}\left(P_{{\rm tx,R}},\ K,\ M\right)}{{P}_{{\rm tot}}\left(P_{{\rm tx,R}},\ K,\ M\right)},\label{APP2_EE}
\end{align}
where
\begin{align}
\widetilde{{\cal R}}\left(P_{{\rm tx,R}},\ K,\ M\right)=K\min\left\{ \widetilde{{\cal R}}_{k}^{(1)},\ \widetilde{{\cal R}}_{k}^{(2)}\right\} ,\label{APP2_R}
\end{align}
with
\begin{align}
\widetilde{{\cal R}}_{k}^{(1)} & \xrightarrow{a.s.}\int_{R_{\min}}^{R_{\max}}\log_{2}\left(1+\frac{\left(M-K\right)P_{{\rm tx,U}}\beta_{k}^{'}}{P_{{\rm tx,d}}{\tilde{A}}_{1}+\sigma_{{\rm R}}^{2}}\right)f\left(l_{k}\right)dl_{k},\label{eq:-10}\\
\widetilde{{\cal R}}_{k}^{(2)} & \xrightarrow{a.s.}\int_{R_{\min}}^{R_{\max}}\log_{2}\left(1+\frac{\left(M-K\right)P_{{\rm tx,R}}}{\left(P_{{\rm tx,R}}{\tilde{\beta}}_{k+K}+\sigma_{{\rm {D}}}^{2}\right)\tilde{{A}}_{2}}\right)\nonumber \\
 & \times f\left(l_{k+K}\right)dl_{k+K},\\
\tilde{{A}}_{1} & =\frac{cK}{2K\rho_{{\rm r}}\left(R_{\max}^{2}-R_{\min}^{2}\right)}\Bigg\{ R_{\max2}^{2}F_{1}\left(1,\frac{1}{\alpha};\frac{\alpha+2}{\alpha};\vphantom{\frac{R_{\max}^{\alpha}}{2Kc\rho_{{\rm r}}}}\right.\nonumber \\
 & \quad\left.-\frac{R_{\max}^{\alpha}}{2Kc\rho_{{\rm r}}}\right)-R_{\min2}^{2}F_{1}\left(1,\frac{1}{\alpha};\frac{\alpha+2}{\alpha};-\frac{R_{\min}^{\alpha}}{2Kc\rho_{{\rm r}}}\right)\Bigg\},\\
\tilde{A}_{2} & =\frac{K}{c\left(R_{\max}^{2}-R_{\min}^{2}\right)}\Bigg\{\frac{1}{2K\rho_{{\rm r}}}\frac{R_{\max}^{2\left(\alpha+1\right)}-R_{\min}^{2\left(\alpha+1\right)}}{c\left(\alpha+1\right)}\nonumber \\
 & \quad+\frac{2\left(R_{\max}^{\alpha+2}-R_{\min}^{\alpha+2}\right)}{\alpha+2}\Bigg\}^{2}.
\end{align}

Again, ${P}_{{\rm tot}}\left(P_{{\rm tx,R}},\ K,\ M\right)$ is given
by \eqref{power_C}.
\end{thm}
\begin{IEEEproof}
See Appendix II.
\end{IEEEproof}
\begin{rem}
\noindent Theorem 2 characterizes the relationship between the EE
and $\left(P_{{\rm tx,R}},\ K,\ M\right)$ under the condition of
equal power allocation and random device locations. According to Theorem
2, we are capable of evaluating the EE without using any channel coefficients
and without complex matrix calculations. This results in a substantial
complexity reduction of the real-time online computation. However,
as far as solving the optimization problem associated with the energy-efficient
resource allocation is concerned, \eqref{APP2_EE} remains excessively
complex due to the tedious integral in \eqref{APP2_R}.
\end{rem}
In order to circumvent the aforementioned obstacle, a lower bound
of \eqref{APP2_EE} is derived as follows. 

\begin{cor}
A lower bound of \eqref{APP2_EE} is given by
\begin{align}
\widetilde{\eta}_{{\rm EE}}\left(P_{{\rm tx,R}},K,M\right) & \geq\widetilde{{\eta}}_{{\rm EELB}}\left(P_{{\rm tx,R}},K,M\right)\label{E_UB}\\
 & =\frac{\left(1-\frac{2K}{T}\right)\frac{BK}{2}\widetilde{{\cal R}}_{{\rm LB}}\left(P_{{\rm tx,R}},K,M\right)}{{P}_{{\rm tot}}\left(P_{{\rm tx,R}},K,M\right)},\nonumber
\end{align}
where 
\begin{align}
\begin{split}\widetilde{{\cal R}}_{{\rm LB}}\left(P_{{\rm tx,R}},K,M\right) & =\min\left\{ \widetilde{{\cal R}}_{{\rm LB}}^{\left(1\right)},\widetilde{{\cal R}}_{{\rm LB}}^{\left(2\right)}\right\} ,\ 
\end{split}
\label{R_LB}
\end{align}
with
\begin{equation}
\begin{split}
{\cal \widetilde{R}}_{{\rm LB}}^{\left(1\right)} & =\log_{2}\left(1+\frac{\left(M-K\right)KP_{{\rm tx,d}}}{\left(P_{{\rm tx,d}}\tilde{A}_{1}+\sigma_{{\rm R}}^{2}\right)\tilde{A}_{2}}\right),\\
{\cal \widetilde{R}}_{{\rm LB}}^{\left(2\right)} & =\log_{2}\left(1+\frac{\left(M-K\right)KP_{{\rm tx,R}}}{\left(P_{{\rm tx,R}}\tilde{A}_{1}+K\sigma_{{\rm D}}^{2}\right)\tilde{A}_{2}}\right).
\end{split}
\end{equation}
\end{cor}
\begin{IEEEproof}
See Appendix III.
\end{IEEEproof}
\begin{rem}
\noindent It can be observed from Corollary 2 that the lower bound
$\widetilde{{\eta}}_{{\rm EELB}}\left(P_{{\rm tx,R}},M,K\right)$
and upper bound $\widetilde{{\eta}}_{{\rm EEUB}}\left(P_{{\rm tx,R}},M,K\right)$
derived are represented by a simple closed-form expression without
the tedious integral in \eqref{APP2_R}, which is significantly beneficial
for efficiently solving the optimization problem associated with our
energy-efficient resource allocation. 
\end{rem}

\section{Energy-efficient resource allocation for maximizing the lower bound
of the EE }

Let us commence with a brief discussion about the rationale and significance
of the analytical results we have obtained so far. Theorem 1 quantifies
the EE of the massive MIMO aided mIoT network considered under the
assumption that the positions of the devices are known. The network
considered in this paper is a narrow-band mIoT (NB-mIoT) network,
which has been standardized in 3GPP Release 13 \cite{[2]} to support
a large number of low-power devices \cite{[3]}. In addition, considering
more general and practical mIoT application scenarios, e.g., environmental
monitoring and agriculture inspection, where the devices are i.u.d.
and their energy is limited, it is significantly vital for controlling
electrical devices to properly address issues related to QoS and energy
distribution. Theorem 2 gives the exact integral expression of the
EE. However, it remains an open challenge to deal with an optimization
problem whose objective function (i.e., the EE herein) is given by
complex integrals. Traditional optimization tools, such as convex
optimization, genetic algorithms, exhaustive search and so forth,
become futile in this scenario. As a remedy, in Corollary 2, a simple
lower bound of the EE is derived, where the tedious integral vanishes.
Naturally, in this section, we reformulate the original energy-efficient
resource allocation problem \eqref{eq_ee_1} as the following optimization
problem\footnote{This optimization problem is quite different from the conventional
resource allocation problem that targets at specific mIoT device pairs.
In this paper, we intend to optimize the EE, which is a system-level
metric to be determined by solving the resource allocation optimization
problem of \eqref{LSF_R_U}. In particular, when solving \eqref{LSF_R_U},
it is unnecessary for the relay to know the SSF and LSF components
of the CSI of each mIoT device pair.} that maximizes the lower bound of the EE.
\begin{align}
\max_{P_{{\rm tx,R}},K,M}\quad & \widetilde{{\eta}}_{{\rm EELB}}\left(P_{{\rm tx,R}},\ K,\ M\right),\nonumber \\
{\rm s.t.\quad} & {{\rm C1^{'}}:}\quad M\in\left\{ 1,\ 2,\dots,\ M_{\max}\right\} ,\nonumber \\
 & {{\rm C2^{'}}:\quad}K\in\left\{ 1,\ 2,\ \dots,M-1\right\} ,\label{LSF_R_U}\\
 & {{\rm C3^{'}}:\quad}0\leq P_{{\rm tx,R}}\leq P_{{\rm R\max}},\nonumber \\
 & {{\rm C4^{'}}:\quad}\widetilde{{\cal R}}_{{\rm LB}}\geq{R}_{0}.\nonumber
\end{align}
It is plausible that \eqref{LSF_R_U} is a non-convex problem, which
remains mathematically challenging to solve. Nonetheless, it has become
tractable. For obtaining the global optimal solution of \eqref{LSF_R_U},
typically we have to carry out brute-force search over the feasible-solution
space, which leads to a potentially prohibitive computational complexity.
Therefore, instead of solving \eqref{LSF_R_U} directly, we propose
a sub-optimal strategy by decomposing \eqref{LSF_R_U} into three
subproblems, i.e., Subproblem I: optimization of the relay's transmit
power $P_{{\rm tx,R}}$ for a given $M,K$; Subproblem II: optimization
of the number of relay antennas $M$ for the given $P_{{\rm tx,R}}$
and $K$; and Subproblem III: optimization of the number of active
mIoT device pairs $K$ for the given $P_{{\rm tx,R}}$ and $M$. Then,
an iterative strategy is used, which solves this pair of subproblems
sequentially in each iteration, as detailed below.

\subsection{Subproblem I: Optimization of the Relay's Transmit Power}

For a given $K$ and $M$, the subproblem of optimizing $P_{{\rm tx,R}}$
is written as
\begin{equation}
\begin{split}\max_{P_{{\rm tx,R}}}\  & \widetilde{\eta}_{{\rm EELB}}\left(P_{{\rm tx,R}}\right),\\
{\rm s.t}.\  & {\rm C3^{'},\ C4'},
\end{split}
\label{jpa}
\end{equation}
which is a non-convex fractional programming problem due to the non-differentiable
objective function $\widetilde{\eta}_{{\rm EELB}}\left(P_{{\rm tx,R}}\right)$.
By introducing a slack variable $\lambda\ (\lambda>0)$, \eqref{jpa}
is transformed into a quasi-convex fractional programming problem
as follows. 
\begin{equation}
\begin{split}\max_{P_{{\rm tx,R}},\lambda}\  & \widebar{\widetilde{\eta}}_{{\rm EELB}}\left(P_{{\rm tx,R}},\lambda\right)\\
{\rm s.t.}\  & {\rm C3}^{'},\ \lambda\geq R_{0},{\cal \widetilde{R}}_{{\rm LB}}^{\left(1\right)}\geq\lambda,\ {\cal \widetilde{R}}_{{\rm L}{\rm B}}^{\left(2\right)}\geq\lambda,
\end{split}
\label{jpa_1}
\end{equation}
where $\widebar{\widetilde{\eta}}_{{\rm EELB}}=\frac{\left(1-\frac{2K}{T}\right)\frac{BK}{2}\lambda}{P_{{\rm CF}}+E_{{\rm PA}}P_{{\rm tx,R}}}$
with $P_{{\rm CF}}=\frac{\eta_{{\rm PA,d}}^{-1}}{2}\left(1-\frac{2K}{T}\right)KP_{{\rm tx,d}}+\frac{4\eta_{{\rm PA,d}}^{-1}}{T}K^{2}\rho_{{\rm r}}\sigma_{{\rm R}}^{2}+P_{{\rm C}}$
and $E_{{\rm PA}}=\frac{\eta_{{\rm PA,R}}^{-1}}{2}\left(1-\frac{2K}{T}\right)$.

The objective of the optimization problem \eqref{jpa_1} is in a quasi-concave
fractional form, which is difficult to address directly. Therefore,
using Dinkelbach's algorithm\cite{dinkelbach1967nonlinear,Yang_R1},
we transform it into a parameterized subtractive form as follows.
\begin{equation}
\begin{split}F\left(\xi\right)\triangleq\max_{P_{{\rm tx,R}},\lambda} & \left(1-\frac{2K}{T}\right)\frac{BK}{2}\lambda\\
 & -\xi\left(P_{{\rm CF}}+\frac{\eta_{{\rm PA,R}}^{-1}}{2}\left(1-\frac{2K}{T}\right)P_{{\rm tx,R}}\right)\\
{\rm s.t.} & \;{\rm C3}^{'},\ \lambda\geq R_{0},{\cal \widetilde{R}}_{{\rm LB}}^{\left(1\right)}\geq\lambda,\ {\cal \widetilde{R}}_{{\rm L}{\rm B}}^{\left(2\right)}\geq\lambda.
\end{split}
\label{jpa_2}
\end{equation}

In \eqref{jpa_2}, $F\left(\xi\right)$ is a strictly decreasing and
continuous function, which is convex for all $\xi\in\mathbb{R}$.
Moreover, $F\left(\xi\right)=0$ has a unique solution denoted by
$\xi^{*}$. We know that $F\left(\xi^{*}\right)$ and the objective
function of \eqref{jpa_1} result in the same optimal solution, and
the optimal objective function value of \eqref{jpa_1} is $\xi^{*}$
\cite{dinkelbach1967nonlinear,Yang_R1}. Therefore, the primal problem
\eqref{jpa_1} is equivalent to the newly defined parametric problem
\eqref{jpa_2}. Let us now turn to solving the problem \eqref{jpa_2}.

It is plausible that all the constraints of \eqref{jpa_2} are either
affine or convex w.r.t $\left(\lambda,P_{{\rm tx,R}}\right)$ for
a given $M$, $K$ and $\xi$. Similarly, the objective function of
\eqref{jpa_2} is also affine w.r.t $\left(\lambda,P_{{\rm tx,R}}\right)$.
As a result, \eqref{jpa_2} is a concave optimization problem. 
It can be readily verified that \eqref{jpa_2} satisfies Slater's
condition \cite{boyd_convex_2004}, hence the optimal solution of
\eqref{jpa_2} may be obtained equivalently by solving its Lagrangian
dual problem
\begin{align}
\min_{\mu_{1},\mu_{2},\mu_{3},\mu_{4}\geq0}\max_{P_{{\rm tx,R}},\lambda} & {\cal L}\left(P_{{\rm tx,R}},\lambda,\mu_{1},\mu_{2},\mu_{3},\mu_{4}\right),\label{dual_1}
\end{align}
where
\begin{equation}
\begin{split} & {\cal L}\left(P_{{\rm tx,R}},\lambda,\mu_{1},\mu_{2},\mu_{3},\mu_{4}\right)=\left(1-\frac{2K}{T}\right)\frac{BK}{2}\lambda\\
 & \;-\xi\left(P_{{\rm CF}}+\frac{\eta_{{\rm PA,R}}^{-1}}{2}\left(1-\frac{2K}{T}\right)\times P_{{\rm tx,R}}\right)+\mu_{1}\left({\cal \widetilde{R}}_{{\rm LB}}^{\left(1\right)}-\lambda\right)\\
 & \;+\mu_{2}\left(\widetilde{{\cal R}}_{{\rm LB}}^{\left(2\right)}-\lambda\right)+\mu_{3}\left(P_{{\rm R}\max}-P_{{\rm tx,R}}\right)+\mu_{4}\left(\lambda-R_{0}\right),
\end{split}
\label{dual_2}
\end{equation}
and $\mu_{1}$, $\mu_{2}$, $\mu_{3}$, $\mu_{4}$ are the Lagrange
multipliers.

The dual problem \eqref{dual_1} can be decomposed into two layers:
the inner-layer maximization problem and outer-layer minimization
problem. The optimal solution of \eqref{dual_1} may be readily obtained
by an iterative method. To elaborate a little further, we first solve
the following inner-layer maximization problem
\begin{align}
\max_{P_{{\rm tx,R}},\lambda} & {\cal L}\left(P_{{\rm tx,R}},\lambda,\mu_{1},\mu_{2},\mu_{3},\mu_{4}\right)
\end{align}
for the fixed Lagrange multipliers $\mu_{1}$, $\mu_{2}$, $\mu_{3}$
and $\mu_{4}$, as well as for the given parameters $\xi$, $M,\ K$.
Let the first-order derivatives of ${\cal L}$ w.r.t. ($\lambda,P_{{\rm tx,R}}$)
be zero, yielding
\begin{equation}
\begin{split} & \frac{\partial{\cal L}}{\partial\lambda}=\left(1-\frac{2K}{T}\right)\frac{BK}{2}-\mu_{1}-\mu_{2}+\mu_{4}=0,\\
 & \frac{\partial{\cal L}}{\partial P_{{\rm tx,R}}}=\frac{\xi\eta_{{\rm PA,R}}^{-1}}{2}\left(\frac{2K}{T}-1\right)-\mu_{3}\\
 & \quad+\frac{\mu_{2}}{\ln2}\left(\frac{\alpha_{1}+1}{\left(\alpha_{1}+1\right)P_{{\rm tx,R}}+\alpha_{2}}-\frac{1}{P_{{\rm tx,R}}+\alpha_{2}}\right)=0,
\end{split}
\end{equation}
where $\alpha_{1}=\frac{\left(M-K\right)K}{\tilde{A}_{2}\tilde{A}_{1}}$
and $\alpha_{2}=\frac{K\sigma_{{\rm D}}^{2}}{\tilde{A}_{1}}$. The
optimal transmit power $P_{{\rm tx,R}}^{*}$ is then calculated as
\begin{equation}
P_{{\rm tx,R}}^{*}=\frac{-\left(\alpha_{1}+2\right)\alpha_{2}+\sqrt{\left(\alpha_{1}+2\right)^{2}\alpha_{2}^{2}+4\left(\alpha_{1}+1\right)\alpha_{3}}}{2\left(\alpha_{1}+1\right)},\label{ocs}
\end{equation}
with $\alpha_{3}=\frac{\alpha_{1}\alpha_{2}\mu_{2}}{\left(\mu_{3}+\frac{\xi\eta_{{\rm PA,R}}^{-1}}{2}\left(1-\frac{2K}{T}\right)\right)\ln2}-\alpha_{2}^{2}$
and
\begin{align}
\mu_{4} & =\left(\frac{2K}{T}-1\right)\frac{BK}{2}+\mu_{1}+\mu_{2}.\label{mu_3}
\end{align}

By substituting \eqref{mu_3} into \eqref{dual_2}, \eqref{dual_1}
is rewritten as follows:
\begin{align}
\min_{\mu_{1},\mu_{2},\mu_{3}\geq0}\max_{P_{{\rm tx,R}}} & \ {\cal \hat{L}}\left(P_{{\rm tx,R}},\mu_{1},\mu_{2},\mu_{3}\right),
\end{align}
where
\begin{align}
 & \hat{{\cal L}}\left(P_{{\rm tx,R}},\mu_{1},\mu_{2},\mu_{3}\right)=\left(1-\frac{2K}{T}\right)\frac{BK}{2}R_{0}\nonumber \\
 & \;-\xi\left(P_{{\rm CF}}+\frac{\eta_{{\rm PA,R}}^{-1}}{2}\left(1-\frac{2K}{T}\right)P_{{\rm tx,R}}\right)+\mu_{1}\left({\cal \widetilde{R}}_{{\rm LB}}^{\left(1\right)}-R_{0}\right)\nonumber \\
 & \;+\mu_{2}\left(\widetilde{{\cal R}}_{{\rm LB}}^{\left(2\right)}-R_{0}\right)+\mu_{3}\left(P_{{\rm R}\max}-P_{{\rm tx,R}}\right).
\end{align}

For the outer-layer minimization problem, since the Lagrange function
$\hat{{\cal L}}$ is differentiable, the gradient method may be readily
used for updating the Lagrange multipliers $\mu_{1},\mu_{2}$ and
$\mu_{3}$ as follows.
\begin{equation}
\begin{split}\mu_{1}^{\left(n+1\right)} & =\left[\mu_{1}^{\left(n\right)}-\tau_{\mu_{1}}\left({\cal \widetilde{R}}_{{\rm LB}}^{\left(1\right)}-R_{0}\right)\right]^{+},\\
\mu_{2}^{\left(n+1\right)} & =\left[\mu_{2}^{\left(n\right)}-\tau_{\mu_{2}}\left({\cal \widetilde{R}}_{{\rm LB}}^{\left(2\right)}\left(P_{{\rm tx,R}}^{*}\right)-R_{0}\right)\right]^{+},\\
\mu_{3}^{\left(n+1\right)} & =\left[\mu_{3}^{\left(n\right)}-\tau_{\mu_{3}}\left(P_{{\rm R}\max}-P_{{\rm tx,R}}^{*}\right)\right]^{+},
\end{split}
\label{ite_mu}
\end{equation}
where the superscript `$n$' denotes the iteration index, $\tau_{\mu_{1}}$,
$\tau_{\mu_{2}}$ and $\tau_{\mu_{3}}$ are the step sizes used for
moving in the direction of the negative gradient for the dual variables
$\mu_{1}$, $\mu_{2}$ and $\mu_{3}$, respectively.

Finally, the optimization problem \eqref{jpa_1} under the given $K$
and $M$ can be solved by a two-stage iterative algorithm. In the
first stage, the parameter $\xi$ is updated using Dinkelbach's method
\cite{dinkelbach1967nonlinear,Yang_R1}. In the second stage, the
optimal Lagrange multipliers and $P_{{\rm tx,R}}^{*}$ are obtained
for the given $\xi$. The detailed iterative procedure is summarized
in Algorithm 1.
\begin{algorithm}[tbh]
\protect\caption{Iterative algorithm for optimizing the transmit power of the relay}

\begin{itemize}
\item \textbf{Initialization}: $\xi^{\left(0\right)}>0$, $\mu_{10}>0$,
$\mu_{20}>0$, $\mu_{30}>0$, $\epsilon_{1}>0$, $\epsilon_{2}>0$,
$\tau_{\mu_{1}}$, $\tau_{\mu_{2}}$, $\tau_{\mu_{3}}$ and $m=0$
\item \textbf{Repeat} (corresponding to the first stage)
\item $m\leftarrow m+1$, $n=0$, $\mu_{1}^{\left(0\right)}=\mu_{10},\ \mu_{2}^{\left(0\right)}=\mu_{20}$,
$\mu_{3}^{\left(0\right)}=\mu_{30}$.
\begin{enumerate}
\item \textbf{Repeat} (corresponding to the second stage)
\item $n\leftarrow n+1$.
\item Calculate $P_{{\rm tx,R}}^{*}$ with \eqref{ocs} under the given
$\mu_{1}^{\left(n-1\right)}$, $\mu_{2}^{\left(n-1\right)}$, $\mu_{3}^{\left(n-1\right)}$
and $\xi^{\left(m-1\right)}$.
\item Update $\mu_{1}^{\left(n\right)}$, $\mu_{2}^{\left(n\right)}$ and
$\mu_{3}^{\left(n\right)}$ by \eqref{ite_mu}.
\item $\Delta\mu_{1}=\mu_{1}^{\left(n\right)}-\mu_{1}^{\left(n-1\right)}$,
$\Delta\mu_{2}=\mu_{2}^{\left(n\right)}-\mu_{2}^{\left(n-1\right)}$,
$\Delta\mu_{3}=\mu_{3}^{\left(n\right)}-\mu_{3}^{\left(n-1\right)}$.
\item \textbf{Until} $\begin{vmatrix}\Delta\mu_{1}\end{vmatrix}\leq\epsilon_{1}$,
$\begin{vmatrix}\Delta\mu_{2}\end{vmatrix}\leq\epsilon_{1}$ and $\begin{vmatrix}\Delta\mu_{3}\end{vmatrix}\leq\epsilon_{1}$.
\item $\xi^{\left(m\right)}=\frac{\left(1-\frac{2K}{T}\right)\frac{BK}{2}\widetilde{{\cal R}}_{{\rm LB}}\left(P_{{\rm tx,R}}^{*}\right)}{P_{{\rm CF}}+\frac{\eta_{{\rm PA,R}}^{-1}}{2}\left(1-\frac{2K}{T}\right)P_{{\rm tx,R}}^{*}}$
\end{enumerate}
\item \textbf{Until}
\begin{align*}
\left|\left(1-\frac{2K}{T}\right)\frac{BK}{2}\widetilde{{\cal R}}_{{\rm LB}}\left(P_{{\rm tx,R}}^{*}\right)-\xi^{\left(m\right)}\right.\\
\left.\times\left(P_{{\rm CF}}+\frac{\eta_{{\rm PA,R}}^{-1}}{2}\left(1-\frac{2K}{T}\right)P_{{\rm tx,R}}^{*}\right)\right| & \leq\epsilon_{2}
\end{align*}
is satisfied.
\item $\xi^{*}\leftarrow\xi^{\left(m\right)}$.
\item \textbf{Output} $\left(\xi^{{\rm *}},\ P_{{\rm tx,R}}^{{\rm *}}\right)$.
\end{itemize}
\end{algorithm}


\subsection{Subproblem II: Optimization of the Number of Relay Antennas}

Given $P_{{\rm tx,R}}$ and $K$, the optimization of the number of
relay antennas is formulated as
\begin{equation}
\begin{split}\max_{M} & \ \widetilde{{\eta}}_{{\rm EELB}}\left(M\right),\\
{\rm s.t.} & \ {\rm C1^{'}},{\rm C4^{'}}.
\end{split}
\label{onaup}
\end{equation}
We can observe that the optimization variable $M$ in \eqref{onaup}
takes integer value. Therefore, this optimization problem is an intractable
non-convex problem. To address this challenge, we firstly relax $M$
to a real variable $M^{'}$. Then, by introducing a slack variable
$\chi\ (\chi>0)$, \eqref{onaup} is transformed into a quasi-convex
fraction programming problem as follows. 
\begin{equation}
\begin{split}\max_{M^{'},\chi}\  & \widehat{\widetilde{\eta}}_{{\rm EELB}}\left(M^{'},\chi\right),\\
{\rm s.t.}\  & {\rm C1}^{'},\ \chi\geq R_{0},{\cal \widetilde{R}}_{{\rm LB}}^{\left(1\right)}\geq\chi,\ {\cal \widetilde{R}}_{{\rm L}{\rm B}}^{\left(2\right)}\geq\chi,
\end{split}
\label{jpa_1b}
\end{equation}
where $\widehat{\widetilde{\eta}}_{{\rm EELB}}=\frac{\left(1-\frac{2K}{T}\right)\frac{BK}{2}\chi}{P_{{\rm fixm}}+P_{{\rm cm}}M^{'}}$
with $P_{{\rm fixm}}=P_{{\rm PA}}+P_{{\rm FIX}}+P_{{\rm PA}}+2KP_{{\rm U}}+P_{{\rm SYN}}+\frac{B}{T}\frac{K^{3}}{3L_{{\rm R}}}$
and $P_{{\rm cm}}=P_{{\rm R}}+\frac{B}{T}\frac{8K^{2}}{L_{{\rm R}}}+B\left(1-\frac{2K}{T}\right)\frac{4K}{L_{{\rm R}}}+\frac{B}{T}\frac{9K^{2}+3K}{3L_{{\rm R}}}$.

Compared to \eqref{jpa_1}, it is easy to see that \eqref{jpa_1b}
has the completely exact same form as \eqref{jpa_1}. Therefore, imitating
Algorithm 1, we can obtain an optimal solution ${M^{'}}^{*}$ of \eqref{jpa_1b}.
Finally, the optimal solution of $M$ is calculated by ${M}^{*}=\lceil{M^{'}}^{*}\rceil$,
where $\lceil\cdot\rceil$ is the ceiling function.


\subsection{Subproblem III: Optimization of the Number of Active Device Pairs}

Given $P_{{\rm tx,R}}$ and $M$, the optimization of the number of
active device pairs is formulated as
\begin{equation}
\begin{split}\max_{K} & \ \widetilde{{\eta}}_{{\rm EELB}}\left(K\right),\\
{\rm s.t.} & \ {\rm C2^{'}},{\rm C4^{'}}.
\end{split}
\label{onaup_K}
\end{equation}
Since the objective function of \eqref{onaup_K} involves the hypergeometric
function, it is challenging to obtain the closed-form expression of
the optimal solutions $K^{*}$. To elaborate a little further, we
cannot relax $K$ as a continuous variable $K^{'}$ for solving \eqref{onaup_K},
since it is difficult to reformulate the hypergeometric function of
\eqref{onaup_K} into a concave or quasi-concave function w.r.t. $K^{'}$.
Fortunately, the feasible region of \eqref{onaup_K} is $\left\{ 1,\ 2,\ \dots,\ M-1\right\} $,
which is discrete and finite. Therefore, we can efficiently solve
\eqref{onaup_K} using a one-dimensional search method.

So far, Subproblem I, Subproblem II and Subproblem III have been solved
sequentially. In a similar fashion, the optimal solutions to the three
subproblems are treated in turn as the initial values of each other
in the remaining iterations, and the optimization problem \eqref{LSF_R_U}
can then be efficiently solved relying on this iterative method. For
the sake of clarity, our energy-efficient resource allocation algorithm
conceived to maximize the lower bound of the EE is summarized as Algorithm
2.

\begin{algorithm}[tbh]
\begin{itemize}
\item \textbf{Input:} The maximum number of iterations $N_{{\rm loop}}$,
the accuracy tolerance $\epsilon$, and an initial value of the vector
$\left(P_{{\rm tx,R}},K,M\right)$
\item \textbf{Output:} $\left(P_{{\rm tx,R}}^{*},K^{*},M^{*}\right)$
\end{itemize}
\begin{enumerate}
\item Assume $n=1$;
\item Update $P_{{\rm tx,R}}^{\left(n\right)}$ by Algorithm 1;
\item Replace $M^{\left(n\right)}$ by the optimal solution obtained from
solving \eqref{onaup} using the similar method updating $P_{{\rm tx,R}}^{\left(n\right)}$;
\item Replace $K^{\left(n\right)}$ by the optimal solution obtained from
solving \eqref{onaup_K} using the one-dimensional search method;
\item $n\leftarrow n+1$, repeat 2), 3) and 4);
\item If
\begin{align*}
\left|\widetilde{{\eta}}_{{\rm EELB}}\left(P_{{\rm tx,R}}^{\left(n\right)},K^{\left(n\right)},M^{\left(n\right)}\right)-\widetilde{{\eta}}_{{\rm EELB}}\right.\\
\left.\times\left(P_{{\rm tx,R}}^{\left(n-1\right)},K^{\left(n-1\right)},M^{\left(n-1\right)}\right)\right| & <\epsilon
\end{align*}
or $n=N_{{\rm loop}}$,\\
 then stop the iteration;
\item $\left(P_{{\rm tx,R}}^{*},K^{*},M^{*}\right)\gets\left(P_{{\rm tx,R}}^{\left(n\right)},K^{\left(n\right)},M^{\left(n\right)}\right)$.
\end{enumerate}
\protect\caption{Energy-efficient resource allocation algorithm for maximizing the
lower bound of the EE}
\end{algorithm}

\section{Convergence and Computational Complexity Analysis}

\begin{figure}
\begin{centering}
\includegraphics[width=7cm]{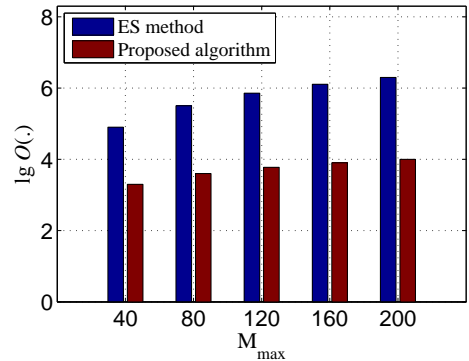}
\par\end{centering}
\centering{}\caption{Complexity comparisons of Algorithm 2 and the ES
method against different $M_{\max}$s. Note that the y-axis uses a
base 10 logarithmic scale.\label{fig:Complex}}
\end{figure}

In this section, we will present a convergence analysis of Algorithm
2 and a detailed complexity analysis so as to get a better insight
into the computational complexity of the proposed algorithm.

From Section V, we know that Algorithm 2 includes three sequential
solving processes corresponding to three subproblems. For Subproblem
I, since \eqref{jpa_2} represents a convex optimization problem w.r.t.
$\left(\lambda,P_{{\rm tx,R}}\right)$, it is guaranteed that the
solutions obtained in the second stage of Algorithm 1 converge to
the optimal solution $P_{{\rm tx,R}}^{*}$ for each given $\xi$.
Moreover, according to \cite{boyd_convex_2004}, the first stage of
Algorithm 1 is also guaranteed to converge, since it sequentially
looks for the optimal value of the univariate parameter $\xi^{{\rm *}}$
with the aid of multiple iterations. Therefore, subproblem I can converge
to a unique value. For Subproblem II, since it has the same solving
process as Subproblem I, Subproblem II can also converge to a fixed
value and obtain an optimal solution of $M$. As for Subproblem III,
the optimal solution of $K$ is obtained by exhaustive search (ES).
As a result, the proposed Algorithm 2 eventually converges. In Section
VII, we have carried out extensive numerical simulations, where the
convergence of Algorithm 2 is always empirically achieved.

From \eqref{eq_ee_1}, we can see that $\eta_{{\rm EE}}$ in the original
optimization problem is a function of the relay power allocation matrix,
${\bf P}$, the set of active mIoT device pairs, ${\cal S}$, and
the number of the relay antennas, $M$. This is a complicated non-convex
problem, so we can only employ the ES method to find its optimal solution.
Because ${\bf P}=\text{diag}\left\{ \sqrt{p_{1}},\dots,\sqrt{p_{K}}\right\} $,
the optimal solution of ${\bf P}$ can be found by searching over
$p_{1},\:p_{2},\:\ldots,\:p_{K}$ under the assumption that each of
them takes discrete values \cite{complexity}. Thus, the computational
complexity of this step is $O(D^{K})$, where $D$ is the number of
power levels that can be taken by $p_{k}$. Since there are $K$ pairs
of active mIoT devices in the set, $S$, and all of them are selected
from the group of all the available sets of active UE pairs ($K\leq M-1$),
the complexity of UE-pair selection is $O(C_{M-1}^{K})$. Furthermore,
$M$ is a discrete and finite variable. As a result, the total complexity
of solving \eqref{eq_ee_1} by using the ES method is $O(\sum_{M=1}^{M_{\mathrm{max}}}\sum_{K=1}^{M-1}MC_{M-1}^{K}D^{K})$$=O\left(\left[M_{\mathrm{max}}D\left(2(D+1)^{M_{\mathrm{max}}}\right.\right.\right.$$\left.-M_{\mathrm{max}}D-D\right)-2(D+1)^{M_{\mathrm{max}}}$$\left.\left.+2\vphantom{\left(M_{\mathrm{max}}\right)}\right]/(2D^{2})\right)$.
As the largest number of the relay antennas $M_{\max}$ is large enough,
the computational complexity is approximately equal to $O(M_{\mathrm{max}}D^{M_{\mathrm{max}}-1})$.

The reformulated optimization problem \eqref{LSF_R_U} is also a non-convex
problem. To obtain the global optimal solution of \eqref{LSF_R_U},
we have to use the ES method over the feasible-solution space. Thus,
the total complexity is $O\left(D'M_{\mathrm{max}}(M_{\mathrm{max}}-1)\right)$,
where $D'$ denotes the number of power level of relay's total transmit
power, $P_{{\rm tx,R}}$. When $M_{\max}$ is large enough, the computational
complexity is approximately equal to $O\left(D'M_{\mathrm{max}}^{2}\right)$.
Compared with the computational complexity of solving \eqref{eq_ee_1},
it is obvious that the computational complexity of solving \eqref{LSF_R_U}
has been significantly reduced.

On the other hand, since the ES method can lead to a prohibitive computational
complexity, we propose Algorithm 2 to solve \eqref{LSF_R_U} by decomposing
this problem into three subproblems. Assume that $I_{\mathrm{in1}}$
is the number of inner iterations required for reaching convergence
of $\xi$ by using the Dinkelbach's method in subproblem I. The complexity
of updating the relay's total transmit power $P_{{\rm tx,R}}$ is
$O(3I_{\mathrm{in1}}I_{\mathrm{ou1}})$, where $I_{\mathrm{ou1}}$
is the number of outer iterations. Similarly, in subproblem II, the
complexity of updating the number of relay antennas $M$ is $O(3I_{\mathrm{in2}}I_{\mathrm{ou2}})$,
where $I_{\mathrm{in2}}$ and $I_{\mathrm{ou2}}$ are the number of
inner and outer iterations in subproblem II, respectively. The one-dimensional
search method used in subproblem III has a complexity of $O(M_{\mathrm{max}}-1)$.
If Algorithm 2 converges after $I_{\mathrm{loop}}$ iterations, the
total complexity is $O\left(I_{\mathrm{loop}}(3I_{\mathrm{in1}}I_{\mathrm{ou1}}+3I_{\mathrm{in2}}I_{\mathrm{ou2}}+M_{\mathrm{max}}-1)\right)$.
When $M_{\max}$ is large enough, the computational complexity is
approximately equal to $O\left(I_{\mathrm{loop}}M_{\mathrm{max}}\right)$.
We can see that the computational complexity of the proposed algorithm
is greatly less than that of the ES method when $D'$ is comparable
to $I_{\mathrm{loop}}$. In Fig. \ref{fig:Complex}, we compare the
computational complexity of the proposed algorithm with that of the
ES method in terms of the largest number of the relay antennas $M_{\max}$,
where $D'=50$ and $I_{\mathrm{loop}}=50$. It is observed that Algorithm
2 exhibits a complexity reduction significantly compared with the
ES method for any $M_{\max}$.

\section{Numerical Results}

In this section, we evaluate the EE performance of the massive MIMO
aided mIoT network and demonstrate the accuracy of our analytical
results via numerical simulations. As a strong candidate for supporting
mIoT communications \cite{[1]}, a small-cell cellular network is
considered, and the simulation parameters are summarized in Table
I. All the simulation parameters in this table are in accordance with
the narrow-band IoT (NB-IoT) network \cite{[2]} which promises to
improve the cellular systems for mIoT by supporting a large number
of IoT devices \cite{[3]}. As part of 3GPP Release 13 \cite{[2]},
NB-IoT has been standardized for mIoT, and the required bandwidth
for NB-IoT is 180 KHz for both uplink and downlink. Our numerical
studies will demonstrate the efficiency of the proposed optimization
strategy, and the impact of several relevant system parameters on
the optimal relay transmit power, on the optimal number of relay antennas
and on the optimal selection of the active mIoT device pairs.

\begin{table}[tbh]
\centering{} \protect \setlength{\abovecaptionskip}{0pt} \setlength{\belowcaptionskip}{10pt}
\caption{Simulation Parameters}
\begin{tabular}{|l|c|}
\hline
\textbf{Parameter}  & \textbf{Value}\tabularnewline
\hline
Reference distance: $R_{\min}$  & 35\ m \tabularnewline
\hline
LSF model: $\beta_{k}=cl_{k}^{-\alpha}${}  & $10^{-0.53}\big/l_{k}^{3.76}$ \tabularnewline
\hline
Transmission bandwidth: $B$  & 20\ MHz\tabularnewline
\hline
Channel coherence bandwidth: $B_{c}$  & 180\ KHz \tabularnewline
\hline
Channel coherence time: $T_{c}$  & 10\ ms \tabularnewline
\hline
Fixed power consumption: $P_{{\rm FIX}}$  & 18\ W \tabularnewline
\hline
Total noise power: $B\sigma^{2}$  & -96\ dBm\tabularnewline
\hline
Computational efficiency at the relay: $L_{{\rm R}}$  & $12.8\ {\rm Gflops\big/W}$\tabularnewline
\hline
PA efficiency at the relay: $\eta_{{\rm PA,R}}$  & 0.39\tabularnewline
\hline
PA efficiency at devices: $\eta_{{\rm PA},{\rm U}}$  & 0.3\tabularnewline
\hline
Circuit power consumption at the relay: $P_{{\rm R}}$  & 1\ W\tabularnewline
\hline
Circuit power consumption at devices: $P_{{\rm d}}$  & 0.1\ W\tabularnewline
\hline
\end{tabular}
\end{table}

\subsection{Accuracy of the Analytical EE}

\begin{figure*}[t]
\begin{centering}
\includegraphics[width=13cm]{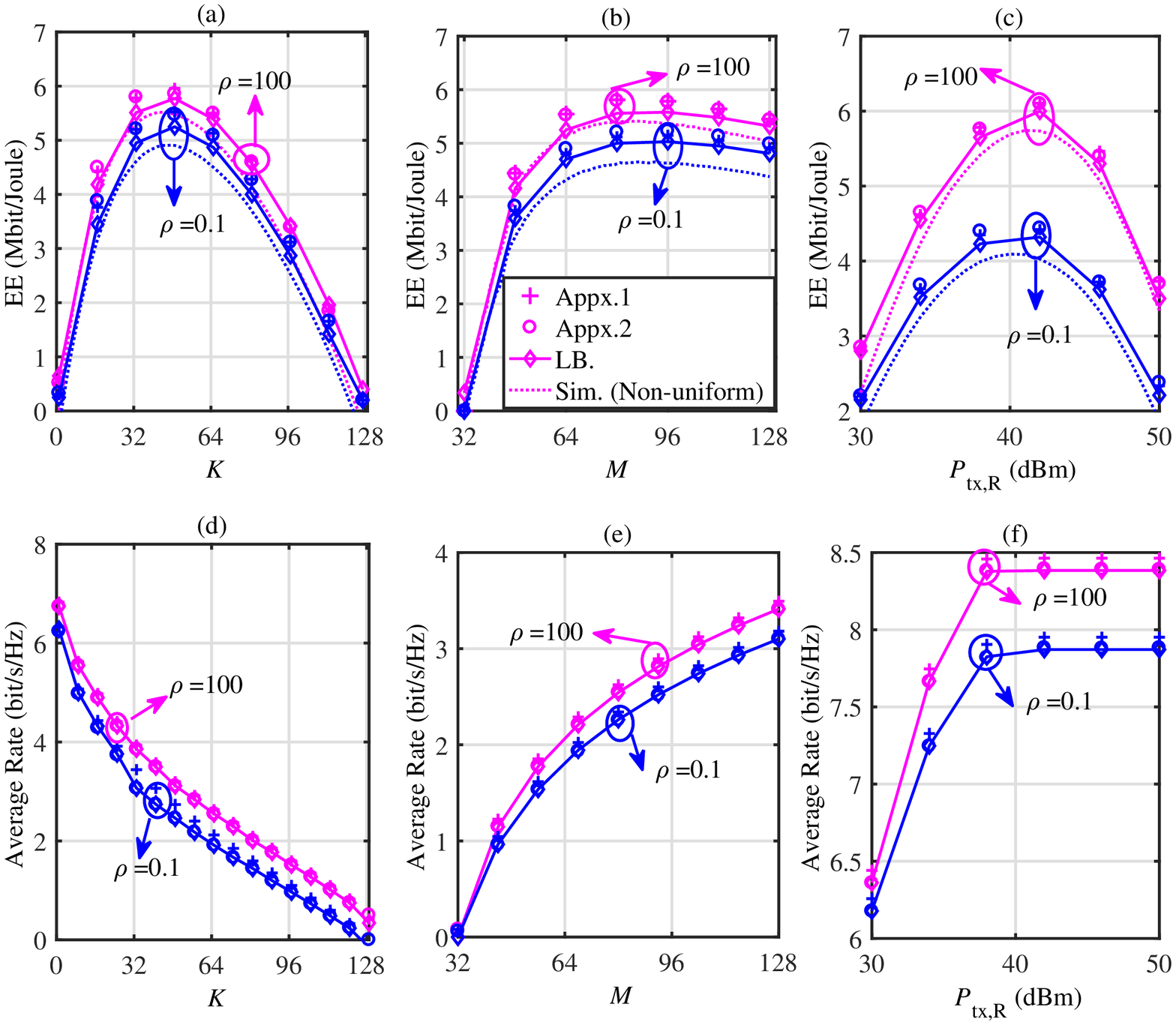}
\par\end{centering}
\protect\caption{The EE and the corresponding average rate versus $K$, $M$ and $P_{{\rm tx,R}}$.
We assume that $M=128$ and $P_{{\rm tx,R}}=30\ {\rm dBm}$ in (a)
and (d), $K=32$ and $P_{{\rm tx,R}}=30\ {\rm dBm}$ in (b) and (e),
while $M=128$ and $K=40$ in (c) and (f). Two scenarios, i.e., $\rho_{{\rm r}}=100$
(indicating that the CE quality is high) and 
 $\rho_{{\rm r}}=0.1$ (indicating that the CE quality is relatively
low), 
 are compared in each sub-figure.}
\label{ee_ar}
\end{figure*}

In Fig. \ref{ee_ar}, the EE ${\eta}_{{\rm EE}}$ and the corresponding
average rate defined as ${\cal R}=\frac{1}{K}\sum_{k=1}^{K}{\cal R}_{k}$
are numerically evaluated assuming $R_{\max}=250\ {\rm m}$, $R_{0}=1\ {\rm bit/s/Hz}$,
and $P_{{\rm tx,d}}=20\ {\rm dBm}$. We also show the analytical EE
given in \eqref{APP1_EE} and the corresponding average rate derived
from \eqref{ar1} (marked as Appx. 1), the analytical EE given in
\eqref{APP2_EE} and the corresponding average rate derived from \eqref{APP2_R}
(marked as Appx. 2), as well as the EE lower bound given in \eqref{E_UB}
and the corresponding rate \eqref{R_LB} (marked as LB). Moreover,
when device locations have non-uniform distribution, simulation values of
the EE (marked as Sim.) are also provided. To construct a non-uniform
distribution across the whole coverage area, we segment the coverage
area into two nested circular cells, where active mIoT devices of the two cells follow uniform distributions with different probability densities. The radii of the nested
circular cells are 100$\,$m and 250$\,$m, respectively.
It can be clearly seen from Fig. \ref{ee_ar} that the analytical
EE expressions derived and the EE lower bound are accurate even in
a system of finite dimensions. Moreover, Fig. \ref{ee_ar}(a)-Fig.
\ref{ee_ar}(c) show the EE versus the number of active mIoT device
pairs, $K$, versus the number of relay antennas $M$, and versus
the relay's transmit power $P_{{\rm tx,R}}$, respectively. Meanwhile,
the corresponding average rate is shown in Fig. \ref{ee_ar}(d)-Fig.
\ref{ee_ar}(f). It can be observed from Fig. \ref{ee_ar}(a)-Fig.
\ref{ee_ar}(c) that given the values of the other parameters, regardless
of whether the quality of CE is high (e.g., when $\rho_{{\rm r}}=100$,
as shown by the purple curves) or not (e.g., when $\rho_{{\rm r}}=0.1$,
as shown by the blue curves), the EE is not a monotonically increasing/decreasing
function of $K$, $M$ or $P_{{\rm tx,R}}$. The optimal value of
$K$, $M$ or $P_{{\rm tx,R}}$ maximizing the EE is usually not on
the boundary. However, the average rate increases with $M$ and $P_{{\rm tx,R}}$,
but decreases with $K$. Explicitly, in order to improve the average
rate of all the active device pairs, the system needs more relay antennas,
higher relay transmit power, or fewer pairs of active mIoT devices
supported simultaneously. Furthermore, as shown in Fig. \ref{ee_ar}(a)-Fig.
\ref{ee_ar}(c), it is worth noting that the lower bound given by
\eqref{E_UB} is close to the analytical EE of \eqref{APP2_EE}. Therefore,
\eqref{E_UB} is a sufficiently tight lower bound. At the same time,
when the assumption that all the devices follow i.u.d. does not hold,
the EE performance will degrade owing to model mismatch.

\subsection{Convergence and Optimality of the Proposed Optimization Strategy}

\begin{figure}[t]
\begin{centering}
\includegraphics[width=7.5cm]{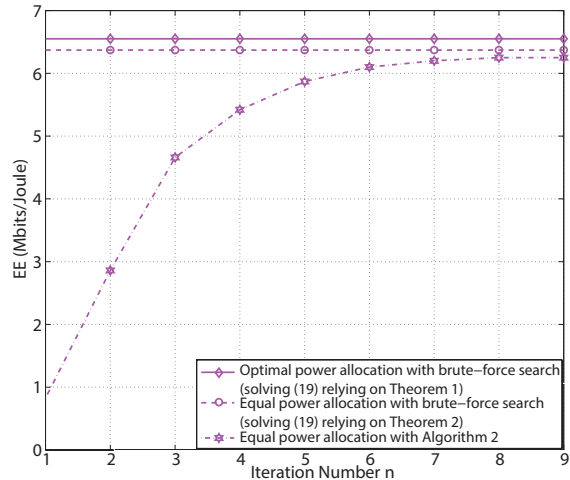}
\par\end{centering}
\protect\caption{Convergence and optimality of the proposed Algorithm 2. Assume ${R}_{0}=1\ {\rm bit/s/Hz}$,
$M_{\max}=128$, $P_{{\rm R\max}}=50\ {\rm dBm}$, $\rho_{{\rm r}}=100$,
and $P_{{\rm tx,U}}=20$ dBm.}
\label{iter}
\end{figure}

In Fig. \ref{iter}, we show the convergence of the proposed iterative
resource allocation approach presented in Algorithm 2 by examining
the EE attained versus the number of iterations. It can be observed
that the maximum EE obtained using Algorithm 2 appears after 8 iterations,
and this maximum EE value is indeed generated by the optimum system
parameters of $\left(P_{{\rm tx,R}}^{*},\ K^{*},M^{*}\right)=\left(36.6,\ 31,\ 81\right)$.
In order to show the optimality of Algorithm 2, we provide a pair
of performance benchmarks that correspond to solving the problem \eqref{eq_ee_1}
based on \textit{Theorem 1} (i.e., using optimal power allocation)
and on \textit{Theorem 2} (i.e., using equal power allocation) via
the high-complexity brute-force searching (i.e., exhaustive searching),
respectively. It is observed that after Algorithm 2 converges, the
gap between the EE values achieved by Algorithm 2 and by solving the
problem \eqref{eq_ee_1} based on \textit{Theorem 2} with the brute-force
searching becomes small. Moreover, the optimum system parameters obtained
by solving \eqref{eq_ee_1} based on \textit{Theorem 2} using the
brute-force searching are $\left(P_{{\rm tx,R}}^{*},\ K^{*},M^{*}\right)=\left(37,\ 30,\ 81\right)$,
which are also close to those achieved by Algorithm 2. Therefore,
the proposed Algorithm 2 is near-optimal. It represents an appealing
design option, because it is capable of substantially reducing the
computational complexity at the expense of a marginal performance
loss. Meanwhile, the gap between the two benchmarking algorithms is
also small, which justifies the employment of the low-complexity equal
power allocation at the relay.

\begin{figure}[t]
\begin{centering}
\includegraphics[width=9cm]{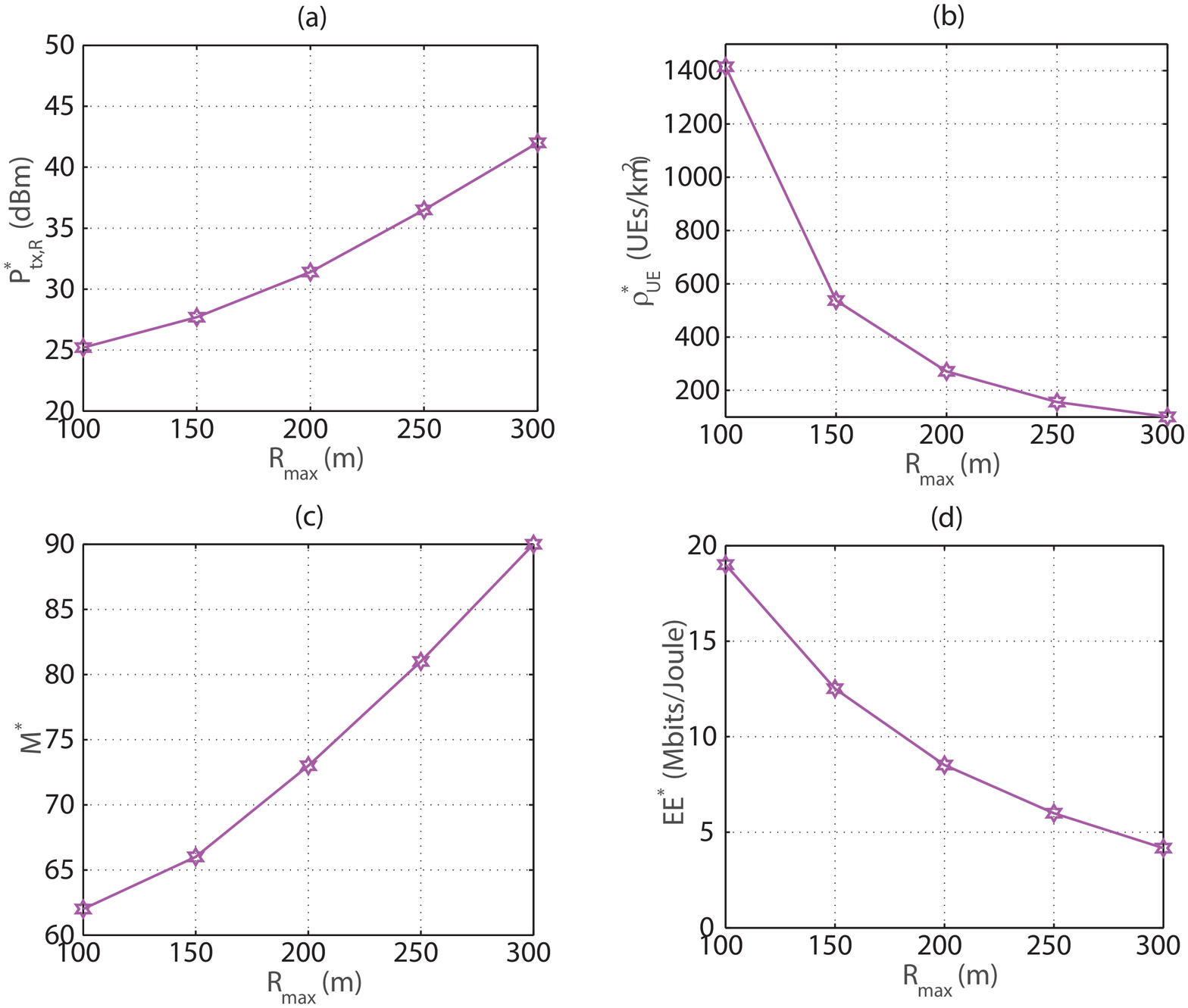}
\par\end{centering}
\protect\caption{The optimum EE and the optimum system parameters $\left(P_{{\rm tx,R}}^{*},\ \rho_{{\rm UE}}^{*},\ M^{*}\right)$
versus different values of $R_{\max}$. Assume $\rho_{{\rm r}}=100$,
${R}_{0}=1\ {\rm bit/s/Hz}$, $M_{\max}=128$, $P_{{\rm R\max}}=50\ {\rm dBm}$,
and $P_{{\rm tx,d}}=20$ dBm.}
\label{Radius}
\end{figure}

In Fig. \ref{Radius}(a)-Fig. \ref{Radius}(c), we investigate the
impact of the coverage area radius $R_{\max}$ on the optimum system
parameters $P_{{\rm tx,R}}^{*},\ \rho_{{\rm UE}}^{*}$, and $M^{*}$,
respectively, where $\rho_{{\rm UE}}=\frac{K}{\pi\left(R_{\max}^{2}-R_{\min}^{2}\right)}$
is the density of the active mIoT device pairs in the given relay's
coverage area. The corresponding optimum EE is shown in Fig. \ref{Radius}(d).
We can readily observe that as $R_{\max}$ becomes large, $P_{{\rm tx,R}}^{*}$
and $M^{*}$ are increased, while $\rho_{{\rm UE}}^{*}$ is decreased.
In other words, for the sake of optimizing the EE, the optimum design
should increase the relay's transmit power, use more antennas at the
relay and reduce the device density, if a larger coverage area of
the relay is required. This conclusion is also supported by the results
shown in Fig. \ref{Radius}(d), where we can see that the optimum
EE is indeed reduced when the coverage area of the relay becomes larger,
provided that the values of the other system parameters remain unchanged.
In fact, it is readily seen that we will have ${\eta}_{{\rm EE}}^{*}\to0$
as $R_{\max}\to+\infty$.

\begin{figure}[t]
\begin{centering}
\includegraphics[width=9cm]{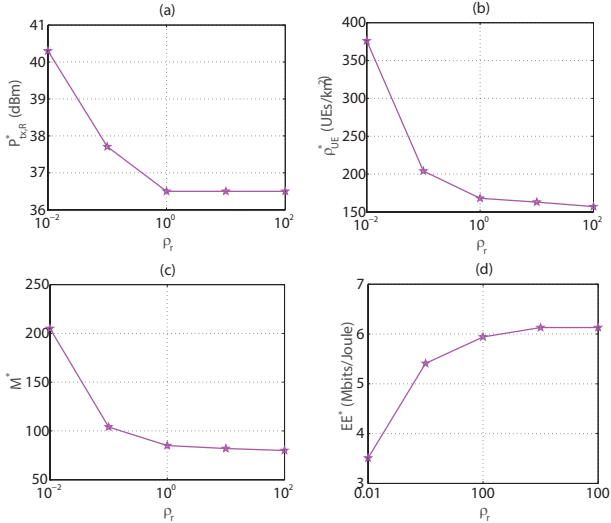}
\par\end{centering}
\protect\caption{The optimum EE and the optimum system parameters $\left(P_{{\rm tx,R}}^{*},\ \rho_{{\rm UE}}^{*},\ M^{*}\right)$
versus the CE quality indicator $\rho_{{\rm r}}$. Assume $R_{\max}=250\ {\rm m}$,
${R}_{0}=1\ {\rm bit/s/Hz}$, $M_{\max}=128$, $P_{{\rm R\max}}=50\ {\rm dBm}$,
and $P_{{\rm tx,d}}=20$ dBm.}
\label{rho_r}
\end{figure}

In Fig. \ref{rho_r}(a)-Fig. \ref{rho_r}(c), we show the impact of
the CE quality indicator $\rho_{r}$ on the optimum system parameters
$\left(P_{{\rm tx,R}}^{*},\ \rho_{{\rm UE}}^{*},\ M^{*}\right)$.
We can observe from these figures that in poor CE scenarios, a higher
relay transmit power, more active devices and more relay antennas
should be used to make the system energy-efficient. Moreover, as expected,
it can be readily observed from Fig. \ref{rho_r}(d) that high-quality
CE is capable of providing a high EE. Additionally, as $\rho_{{\rm r}}$
becomes large, the increase of the EE slows down and converges to
the value that relies on perfect CSI estimation. This implies that
although the system associated with high-quality CE (i.e. $\rho_{{\rm r}}=10^{2}$)
is capable of achieving a better EE performance than the system with
poor CE (i.e. $\rho_{{\rm r}}=10^{-1}$), the pursuit of extremely
high CE accuracy is unnecessary (e.g., $\rho_{{\rm r}}\geq1$ is shown
to be adequate by our simulations).

\begin{figure}[t]
\begin{centering}
\includegraphics[width=9cm]{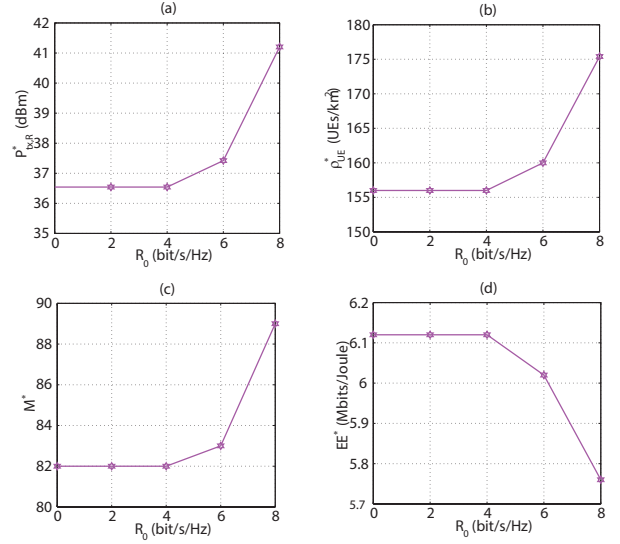}
\par\end{centering}
\protect\caption{The optimum EE and the optimum system parameters $\left(P_{{\rm tx,R}}^{*},\ \rho_{{\rm UE}}^{*},\ M^{*}\right)$
versus the QoS constraint ${R}_{0}$, assuming $R_{\max}=250\ {\rm m}$,
$\rho_{{\rm r}}=100$, $M_{\max}=128$, $P_{{\rm R\max}}=50\ {\rm dBm}$,
and $P_{{\rm tx,d}}=20\ {\rm dBm}$.}
\label{gamma_0}
\end{figure}

In Fig. \ref{gamma_0}(a)-Fig. \ref{gamma_0}(d), we show the optimum
system parameters $\left(P_{{\rm tx,R}}^{*},\ \rho_{{\rm UE}}^{*},\ M^{*}\right)$
and the corresponding optimum EE versus the variations of the QoS
constraint ${R}_{0}$, respectively. It can be observed that for ${R}_{0}\leq4\ {\rm bit/s/Hz}$,
the value of $\left(P_{{\rm tx,R}}^{*},\ \rho_{{\rm UE}}^{*},\ M^{*}\right)$
remains unchanged, but for ${R}_{0}>4\ {\rm bit/s/Hz}$, $\left(P_{{\rm tx,R}}^{*},\ \rho_{{\rm UE}}^{*},\ M^{*}\right)$
increases with the increase of ${R}_{0}$. To be more specific, in
our simulations, the optimum system parameters obtained using Algorithm
2 under the assumption of ${R}_{0}\leq4\ {\rm bit/s/Hz}$ is $\left(P_{{\rm tx,R}}^{*},\ \rho_{{\rm UE}}^{*},\ M^{*}\right)=\left(36.4,155.8,81\right)$,
which results in a QoS constraint of ${\widetilde{{\cal R}}}_{{\rm LB}}=5.53\ {\rm bit/s/Hz}$.
Therefore, in the case of ${R}_{0}\geq5.53\ {\rm bit/s/Hz}$, the
optimal solution is found along the edges of the feasible region that
is affected by the QoS constraint.

\section{Conclusion}

We have provided a series of analytical EE expressions for the mIoT
network using a massive MIMO aided multi-pair DF relay, and proposed
an iterative optimization strategy to maximize the lower bound of
the EE. Firstly, upon assuming that the mIoT device locations are
known \textit{a priori}, a closed-form expression of the EE was derived.
The expression obtained only depends on the LSF channel coefficients
and the configurable system parameters. Secondly, an exact integral
expression of the EE was derived for a more general scenario, where
each device's position is assumed to be an i.u.d random variable in
the relay's coverage area. Moreover, in order to bypass solving complex
integrals, we derived a simple but efficient lower bound of the EE.
Finally, a low-complexity iterative resource allocation strategy was
proposed to maximize this lower bound. Our numerical results demonstrated
the accuracy of the analytical expressions derived, and verified the
effectiveness and convergence speed of the proposed strategy. For
future work, it would be interesting to study the energy-efficient
problems in mobile IoT environments instead of the assumption that
the locations of all the devices are fixed in this paper. We will
also derive an exact integral-based expression of the EE under the
assumption that the locations of devices are non-uniform distribution.

\section*{Appendix I \protect \protect \protect \protect \protect \protect
\protect \\
 Proof of Theorem 1}

Firstly, $\gamma_{k}^{\left(1\right)}$ is derived. Observing \eqref{SNR1},
we have to calculate $\mathbb{E}\left[{\bf f}_{1,k}^{H}{\bf g}_{{\rm S},k}\right]$,
$\mathbb{V}{\rm ar}\left({\bf f}_{1,k}^{H}{\bf g}_{{\rm S},k}\right)$,
${\mathbb{E}}\left[\begin{Vmatrix}{\bf f}_{1,k}\end{Vmatrix}^{2}\right]$
and $\sum_{j\neq k}^{K}\mathbb{E}\left[\begin{vmatrix}{\bf f}_{1,k}^{H}{\bf g}_{{\rm S},j}\end{vmatrix}^{2}\right]$.
Since ${\bf F}_{1}=\left(\hat{{\bf G}}_{{\rm S}}^{H}\hat{{\bf G}}_{{\rm S}}\right)^{-1}\hat{{\bf G}}_{{\rm S}}^{H}$,
we have
\begin{align}
{\bf F}_{1}{\bf G}_{{\rm S}} & ={\bf F}_{1}\left(\hat{{\bf G}}_{{\rm S}}+\tilde{{\bf G}}_{{\rm S}}\right)={\bf I}_{K}+{\bf F}_{1}\tilde{{\bf G}}_{{\rm S}},\label{gg1}
\end{align}
which leads to
\begin{align}
\mathbb{E}\left[{\bf f}_{1,k}^{H}{\bf g}_{{\rm s},k}\right]=1,\mathbb{V}{\rm ar}\left[{\bf f}_{1,k}^{H}{\bf g}_{{\rm S},k}\right]={\mathbb{E}}\left[\begin{vmatrix}{\bf f}_{1,k}^{H}{\tilde{{\bf g}}}_{{\rm S},k}\end{vmatrix}^{2}\right].\label{ICI0}
\end{align}
Then, we calculate $\mathbb{E}\left[\begin{Vmatrix}{\bf f}_{1,k}\end{Vmatrix}^{2}\right]$
and $\mathbb{V}{\rm ar}\left[{\bf f}_{1,k}^{H}{\bf g}_{{\rm S},k}\right]$.
Applying \cite[Theorem 14.3]{couillet_random_2011}, we can obtain
\begin{align}
\begin{Vmatrix}{\bf f}_{1,k}\end{Vmatrix}^{2} & \xrightarrow{a.s.}\frac{\psi}{\frac{M}{K}\phi^{2}-\psi}\frac{1}{K}{\beta}_{k}^{'-1},\label{f1k_V}
\end{align}
where $\phi$ and $\psi$ are the unique solutions of
\begin{align}
\begin{split}\phi=\frac{1}{M}{\rm tr}\left({\bf I}_{M}+\frac{K}{M}\frac{1}{\phi}{\bf I}_{M}\right)^{-1},\\
\psi=\frac{1}{M}{\rm tr}\left({\bf I}_{M}+\frac{K}{M}\frac{1}{\phi}{\bf I}_{M}\right)^{-2}.\label{phi_{1}}
\end{split}
\end{align}
By solving the equation set \eqref{phi_=00003D00003D00003D00007B1=00003D00003D00003D00007D},
we can get $\phi$ and $\psi$. Thus, for \eqref{f1k_V}, we have
\begin{align}
\begin{Vmatrix}{\bf f}_{1,k}\end{Vmatrix}^{2} & \xrightarrow{a.s.}\frac{{\beta}_{k}^{'-1}}{M-K}.\label{f1k_V2}
\end{align}
By the dominated convergence theorem \cite{Pro_Mea} and the continuous
mapping theorem \cite{Asy_Sta}, it is straightforward to show that
\begin{align}
{\mathbb{E}}\left[\begin{Vmatrix}{\bf f}_{1,k}\end{Vmatrix}^{2}\right]- & \frac{{\beta}_{k}^{'-1}}{M-K}\xrightarrow{a.s.}0.\label{f1k_V3}
\end{align}
Furthermore, since ${\bf f}_{{\rm 1},k}$ and $\tilde{{\bf g}}_{{\rm S},k}$
are independent, we have
\begin{align}
\mathbb{V}{\rm ar}\left[{\bf f}_{1,k}^{H}{\bf g}_{{\rm S},k}\right] & ={\mathbb{E}}\left[\begin{vmatrix}{\bf f}_{1,k}^{H}{\tilde{{\bf g}}}_{{\rm S},k}\end{vmatrix}^{2}\right]=\left({\beta}_{k}-{\beta}_{k}^{'}\right){\mathbb{E}}\left[\begin{Vmatrix}{\bf f}_{1,k}\end{Vmatrix}^{2}\right]\nonumber \\
 & \xrightarrow{a.s.}\frac{\left({\beta}_{k}-{\beta}_{k}^{'}\right){\beta}_{k}^{'-1}}{M-K}.\label{VV}
\end{align}

Next, we calculate $\sum_{j\neq k}^{K}{\mathbb{E}}\left[\begin{vmatrix}{\bf f}_{1,k}^{H}\tilde{{\bf g}}_{{\rm S},j}\end{vmatrix}^{2}\right]$.
From \eqref{gg1}, we see that ${\bf f}_{1k}^{H}{\bf g}_{{\rm s},j}={\bf f}_{1,k}^{H}{\tilde{{\bf g}}}_{{\rm S},j}$
for $j\neq k$. 
Thus, we have
\begin{align}
\mathbb{E}\left[\begin{vmatrix}{\bf f}_{1,k}^{H}{\bf g}_{{\rm S},j}\end{vmatrix}^{2}\right] & ={\mathbb{E}}\left[\begin{vmatrix}{\bf f}_{1,k}^{H}{\tilde{{\bf g}}}_{{\rm S},j}\end{vmatrix}^{2}\right]=\left({\beta}_{j}-{\beta}_{j}^{'}\right){\mathbb{E}}\left[\begin{Vmatrix}{\bf f}_{1,k}\end{Vmatrix}^{2}\right]\nonumber \\
 & \xrightarrow{a.s.}\frac{\left({\beta}_{j}-{\beta}_{j}^{'}\right){\beta}_{k}^{'-1}}{M-K}.
\end{align}
Naturally, we can obtain
\begin{align}
\sum_{j\neq k}^{K}{\mathbb{E}}\left[\begin{vmatrix}{\bf f}_{1,k}^{H}\tilde{{\bf g}}_{{\rm S},i}\end{vmatrix}^{2}\right] & \xrightarrow{a.s.}\frac{\beta_{k}^{'-1}\sum_{j\neq k}^{K}\left(\beta_{j}-\beta_{j}^{'}\right)}{M-K}.\label{g111}
\end{align}
Substituting \eqref{ICI0}, \eqref{f1k_V3}, \eqref{VV} and \eqref{g111}
into \eqref{SNR1}, we have
\begin{align}
\gamma_{k}^{\left(1\right)} & \xrightarrow{a.s}\frac{P_{{\rm tx,U}}\left(M-K\right)\beta_{k}^{'}}{P_{{\rm tx,U}}\sum_{i=1}^{K}\left(\beta_{i}-\beta_{i}^{'}\right)+\sigma_{{\rm R}}^{2}}.\label{SNR11}
\end{align}

Secondly, $P_{{\rm tx,R}}$ is derived. According to \eqref{eq:-7}
and \eqref{RPfactor}, we have
\begin{align}
\begin{split} & \label{PRR}P_{{\rm tx,R}}\triangleq{\rm tr}\left({\bf \underline{x}}_{{\rm R}}\underline{{\bf x}}_{{\rm R}}^{H}\right)={\rm tr}\left({\bf F}_{2}{\bf P}\hat{{\bf s}}\hat{{\bf s}}^{H}{\bf P}^{H}{\bf F}_{2}^{H}\right)\\
 & \quad={\rm tr}\left({\bf P}^{2}\left(\hat{{\bf G}}_{{\rm D}}^{H}\hat{{\bf G}}_{{\rm D}}\right)^{-1}\right)=\sum_{k=1}^{K}p_{k}\begin{Vmatrix}{\bf f}_{2,k}\end{Vmatrix}^{2}\\
 & \quad\xrightarrow{a.s.}
\frac{\sum_{k=1}^{K}p_{k}\beta_{k+K}^{'-1}}{M-K}.
\end{split}
\end{align}

Thirdly, $\gamma_{k}^{\left(2\right)}$ is derived. Observing \eqref{SNR2},
we need to calculate $\mathbb{E}\left[{\bf g}_{{\rm D},k}^{H}{\bf f}_{2,k}\right]$,
$\mathbb{V}{\rm ar}\left({\bf g}_{{\rm D},k}^{H}{\bf f}_{2,k}\right)$
and $\mathbb{E}\left[\begin{vmatrix}{\bf g}_{{\rm D},k}^{H}{\bf f}_{2,j}\end{vmatrix}^{2}\right]$.
Following the same methodology used for calculating $\gamma_{k}^{(1)}$,
we have
\begin{equation}
\begin{split} & \mathcal{E}\left[{\bf g}_{{\rm D},k}^{H}{\bf f}_{2,k}\right]\xrightarrow{a.s.}1,\\
 & \mathbb{V}{\rm ar}\left({\bf g}_{{\rm D},k}^{H}{\bf f}_{2,k}\right)\xrightarrow{a.s.}\frac{\left({\beta}_{k+K}-{\beta}_{k+K}^{'}\right){\beta}_{k+K}^{'-1}}{M-K},\\
 & \sum_{j\neq k}^{K}p_{j}\mathbb{E}\left[\begin{vmatrix}{\bf g}_{{\rm D},k}^{H}{\bf f}_{2,j}\end{vmatrix}^{2}\right]\xrightarrow{a.s.}\frac{\left({\beta}_{k+K}-{\beta}_{k+K}^{'}\right)\sum_{j\neq k}^{K}p_{j}{\beta}_{k+K}^{'-1}}{M-K}.
\end{split}
\label{ICI3}
\end{equation}
Substituting \eqref{PRR} and \eqref{ICI3} into \eqref{SNR2}, we
obtain
\begin{align}
\gamma_{k}^{\left(2\right)} & \xrightarrow{a.s.}\frac{p_{k}}{P_{{\rm tx,R}}\left(\beta_{k+K}-\beta_{k+K}^{'}\right)+\sigma_{{\rm D}}^{2}}.
\end{align}
Finally, according to \eqref{ar_e2e} and \eqref{Sim_EE}, we prove
that \eqref{AEE} holds.

\qed 


\section*{Appendix II \protect \protect \protect \protect \protect \protect
\protect \\
 Proof of Theorem 2}

According to the assumption that all the mIoT devices are i.u.d.,
and the strong law of large numbers, the value of ${A}_{1}$ converges
\textit{almost surely} to its expected value when $K$ is sufficiently
large, i.e., we have ${A}_{1}\xrightarrow{a.s.}\tilde{{A}}_{1}$ when
$K\to\infty$. This condition can be easily satisfied in massive MIMO
systems. As a result, $\tilde{A}_{1}$ is calculated as
\begin{align}
 & {A}_{1}\xrightarrow{a.s.}\tilde{{A}}_{1}=K{\mathbb{E}}_{l_{k}}\left[{\tilde{\beta}}_{k}\right]=Kc\int_{R_{\min}}^{R_{\max}}\frac{1}{l_{k}^{\alpha}+2cK\rho_{{\rm r}}}f\left(l_{k}\right)dl_{k}\nonumber \\
 & =\frac{cK}{2K\rho_{{\rm r}}\left(R_{\max}^{2}-R_{\min}^{2}\right)}\Bigg\{ R_{\max2}^{2}F_{1}\left(1,\frac{1}{\alpha};\frac{\alpha+2}{\alpha};-\frac{R_{\max}^{\alpha}}{2Kc\rho_{{\rm r}}}\right)\nonumber \\
 & \;-R_{\min2}^{2}F_{1}\left(1,\frac{1}{\alpha};\frac{\alpha+2}{\alpha};-\frac{R_{\min}^{\alpha}}{2Kc\rho_{{\rm r}}}\right)\Bigg\}.\label{AB2}
\end{align}
Similarly, $\tilde{{A}}_{2}$ is expressed as
\begin{align}
 & {A}_{2}\xrightarrow{a.s.}\tilde{{A}}_{2}=K{\mathbb{E}}_{l_{k+K}}\left[\left(\beta_{k+K}^{'}\right)^{-1}\right]=\frac{K}{c}\int_{R_{\min}}^{R_{\max}}\left(\frac{1}{2K\rho_{{\rm r}}c}\right.\nonumber \\
 & \;\left.\times l_{k}^{2\alpha}+\vphantom{\frac{1}{2K\rho_{{\rm r}}c}}l_{k}^{\alpha}\right)f\left(l_{k}\right)dl_{k}=\frac{K}{c\left(R_{\max}^{2}-R_{\min}^{2}\right)}\nonumber \\
 & \;\times\Bigg\{\frac{1}{2K\rho_{{\rm r}}}\frac{R_{\max}^{2\alpha+2}-R_{\min}^{2\alpha+2}}{c\left(\alpha+1\right)}+\frac{2\left(R_{\max}^{\alpha+2}-R_{\min}^{\alpha+2}\right)}{\alpha+2}\Bigg\}^{2}.\label{AB3}
\end{align}

Next, we will formulate $\widetilde{{\cal R}}\left(P_{{\rm tx,R}},\ K,\ M\right)$
relying on the LSF channel coefficients. According to \eqref{AB2}
and \eqref{AB3}, we have
\begin{align}
\begin{split} & {\cal \widetilde{R}}\left(P_{{\rm tx,R}},\ K,\ M\right)=\sum_{k=1}^{K}\min\Bigg\{{\mathbb{E}}_{l_{1},\dots,l_{K}}\left[{\widebar{\cal R}}_{k}^{(1)}\right],\\
 & \quad\quad{\mathbb{E}}_{l_{1+K},\dots,l_{2K}}\left[{\widebar{\cal R}}_{k}^{(2)}\right]\Bigg\}=K\min\left\{ \widetilde{{\cal R}}_{k}^{(1)},\widetilde{{\cal R}}_{k}^{(2)}\right\} ,
\end{split}
\label{eq:-5}
\end{align}
where 
\begin{align}
 & \widetilde{{\cal R}}_{k}^{(1)}={\mathbb{E}}_{l_{k}}\left[\log_{2}\left(1+\frac{\left(M-K\right)P_{{\rm tx,d}}\beta_{k}^{'}}{P_{{\rm tx,U}}{\widetilde{A}}_{1}+\sigma_{{\rm R}}^{2}}\right)\right]\nonumber \\
 & \quad=\int_{R_{\min}}^{R_{\max}}\log_{2}\left(1+\frac{\left(M-K\right)P_{{\rm tx,d}}\beta_{k}^{'}}{P_{{\rm tx,U}}{\widetilde{A}}_{1}+\sigma_{{\rm R}}^{2}}\right)f\left(l_{k}\right)dl_{k},\nonumber \\
 & \widetilde{{\cal R}}_{k}^{(2)}={\mathbb{E}}_{l_{k+K}}\left[\log_{2}\left(1+\frac{\left(M-K\right)P_{{\rm tx,R}}}{\left(P_{{\rm tx,R}}{\tilde{\beta}}_{k+K}+\sigma_{{\rm D}}^{2}\right)\tilde{{A}}_{2}}\right)\right]\nonumber \\
 & \quad=\int_{R_{\min}}^{R_{\max}}\log_{2}\left(1+\frac{\left(M-K\right)P_{{\rm tx,R}}}{\left(P_{{\rm tx,R}}{\tilde{\beta}}_{k+K}+\sigma_{{\rm D}}^{2}\right)\tilde{{A}}_{2}}\right)\nonumber \\
 & \quad\quad\times f\left(l_{k+K}\right)dl_{k+K}.
\end{align}
\qed


\section*{Appendix III \protect \protect \protect \protect \protect \protect
\protect \\
 Proof of Corollary 2}


According to the convexity of $\log_{2}\left(1+\frac{1}{x}\right)$
and using Jensen's inequality, we obtain the following lower bound:
\begin{align}
\widetilde{{\cal R}}_{k}^{\left(1\right)} & =\mathbb{E}_{l_{k}}\left[\log_{2}\left(1+\frac{\left(M-K\right)P_{{\rm tx,d}}\beta_{k}^{'}}{P_{{\rm tx,d}}{\widetilde{A}}_{1}+\sigma_{{\rm R}}^{2}}\right)\right]\nonumber \\
 & \geq\log_{2}\left(1+\frac{\left(M-K\right)P_{{\rm tx,d}}}{P_{{\rm tx,d}}\tilde{A}_{1}+\sigma_{{\rm R}}^{2}}\left(\mathbb{E}_{l_{k}}\left[\left(\beta_{k}^{'}\right)^{-1}\right]\right)^{-1}\right)\nonumber \\
 & \overset{\left(a\right)}{=}\log_{2}\left(1+\frac{\left(M-K\right)KP_{{\rm tx,d}}}{\left(P_{{\rm tx,d}}\tilde{A}_{1}+\sigma_{{\rm R}}^{2}\right)\tilde{A}_{2}}\right)\triangleq\widetilde{{\cal R}}_{{\rm LB}}^{\left(1\right)},\label{LB1}
\end{align}
where $\left(a\right)$ is obtained by applying
\begin{align}
\mathbb{E}_{l_{k}}\left[\left(\beta_{k}^{'}\right)^{-1}\right] & =\mathbb{E}_{l_{k+K}}\left[\left(\beta_{k+K}^{'}\right)^{-1}\right]=\frac{\tilde{A}_{2}}{K}.
\end{align}
Similarly, we have
\begin{align}
 & \widetilde{{\cal R}}_{k}^{\left(2\right)}=\mathbb{E}_{l_{k}}\left[\log_{2}\left(1+\frac{\left(M-K\right)P_{{\rm tx,R}}}{\left(P_{{\rm tx,R}}{\tilde{\beta}}_{k+K}+\sigma_{{\rm D}}^{2}\right)\tilde{{A}}_{2}}\right)\right]\nonumber \\
 & 
\;\geq\log_{2}\left(1+\frac{\left(M-K\right)P_{{\rm tx,R}}}{\tilde{A}_{2}}\left(P_{{\rm tx,R}}\mathbb{E}_{l_{k+K}}\left[\tilde{\beta}_{k+K}\right]+\sigma_{{\rm D}}^{2}\right)^{-1}\right)\nonumber \\
 & \overset{\left(b\right)}{=}\log_{2}\left(1+\frac{\left(M-K\right)KP_{{\rm tx,R}}}{\left(P_{{\rm tx,R}}\tilde{A}_{1}+K\sigma_{{\rm D}}^{2}\right)\tilde{A}_{2}}\right)\triangleq\widetilde{{\cal R}}_{{\rm LB}}^{\left(2\right)},\label{LB2}
\end{align}
where $\left(b\right)$ is obtained by applying
\begin{align}
\mathbb{E}_{l_{k+K}}\left[\tilde{\beta}_{k+K}\right] & =\mathbb{E}_{l_{k}}\left[\tilde{\beta}_{k}\right]=\frac{\tilde{A}_{1}}{K}.
\end{align}
Substituting \eqref{LB1} and \eqref{LB2} into \eqref{eq:-5}, we
can readily obtain
\begin{align}
\widetilde{{\cal R}}\left(P_{{\rm tx,R}},K,M\right) & \geq K\min\left\{ \widetilde{{\cal R}}_{{\rm LB}}^{\left(1\right)},\widetilde{{\cal R}}_{{\rm LB}}^{\left(2\right)}\right\} \nonumber \\
 & =K\widetilde{{\cal R}}_{{\rm LB}}\left(P_{{\rm tx,R}},K,M\right).
\end{align}
According to \eqref{APP2_EE}, we have
\begin{align}
 & \widetilde{\eta}_{{\rm EE}}\left(P_{{\rm tx,R}},K,M\right)\geq\widetilde{\eta}_{{\rm EELB}}\left(P_{{\rm tx,R}},K,M\right)\nonumber \\
 & \quad=\frac{\left(1-\frac{2K}{T}\right)\frac{BK}{2}\widetilde{{\cal R}}_{{\rm LB}}\left(P_{{\rm tx,R}},K,M\right)}{P_{{\rm tot}}\left(P_{{\rm tx,R}},K,M\right)}.
\end{align}
\qed

\bibliographystyle{IEEEtran}
\bibliography{LSAS_RELAY}

\end{document}